\documentclass[%
reprint,
showpacs,preprintnumbers,
amsmath,amssymb,aps,prd
]{revtex4-1}

\usepackage{graphicx}
\usepackage{dcolumn}
\usepackage{bm}

\begin{document}


\def\be{\begin{equation}}
\def\ee{\end{equation}}
\def\beq{\begin{eqnarray}}
\def\eeq{\end{eqnarray}}

\def\H{\widehat}
\def\n{\widetilde{\nabla}}
\def\w{\widetilde{\delta}}

\def\N{\nonumber}

\def\K{\mathcal{K}}
\def\H{\mathcal{H}}
\def\V{\mathcal{V}}
\def\F{\bar{\Phi}}
\def\T{\bar{\Theta}}

\def\q{\dot{q}}
\def\qq{\ddot{q}}
\def\M{(M^{-1})}
\def\MM{(\mathcal{M}^{-1})}
\def\t{\bar{t}}
\def\bq{\bar{q}}




\title{Hamilton-Jacobi approach for linearly acceleration-dependent Lagrangians}


\author{Alejandro Aguilar-Salas}
\email{alejanaguilar@uv.mx}

\affiliation{Facultad de Matem\'aticas, Universidad Veracruzana,
 Cto. Gonz\'alo Aguirre Beltr\'an s/n, Xalapa, Veracruz 91000, 
M\'exico}

\author{Efra\'\i n Rojas}
\email{efrojas@uv.mx}
\affiliation{Facultad de F\'\i sica, Universidad Veracruzana, 
Cto. Gonz\'alo Aguirre Beltr\'an s/n, Xalapa, Veracruz 91000, 
M\'exico
}%






\begin{abstract}
We develop a constructive procedure for arriving at the 
Hamilton-Jacobi framework for the so-called affine in acceleration 
theories by analysing the canonical constraint structure. 
We find two scenarios in dependence of the order of the emerging 
equations of motion. By properly defining generalized brackets, the 
non-involutive constraints that originally arose, in both scenarios, may be removed so that the resulting involutive Hamiltonian constraints ensure integrability of the theories and, at the same time, lead to the right dynamics in the reduced phase space. In particular, when we have second-order in derivatives equations of motion we are able to detect the gauge invariant sector of the theory by using a suitable approach based on the projection of the Hamiltonians onto the tangential and normal directions of the congruence of curves in the configuration space. Regarding this, we also explore the generators of canonical and gauge transformations of these theories. Further, we briefly outline how to determine the Hamilton principal function $S$ for some particular setups. We apply our findings to some representative theories: a Chern-Simons-like theory in $(2+1)$-dim, an harmonic oscillator in $2D$ and, the geodetic brane cosmology emerging in the context of extra dimensions.

\end{abstract}


\maketitle


\section{\label{sec:intro}Introduction}

There is a broad class of relevant Lagrangian functions that are 
not built from the standard prescription $L= T- V$. Closely related 
to this set are the second-order derivative theories that in many 
cases are considered extensions, or corrections, to the usual physical 
theories, possessing an intricate gauge freedom which is  responsible 
for not allowing the highest derivatives to be solved algebraically 
in the equations of motion. In this sense, particular attention has 
been focused on the so-called \textit{affine in acceleration 
theories}~\cite{Carinena2003,Udriste2011,Rojas2016}, mainly motivated 
by model building associated with some relativistic acceleration 
phenomena including general relativity, electromagnetism, or 
modified gravity theories. These are characterized by a linear dependence 
on the accelerations of the systems and are necessarily singular 
giving rise in general to both gauge ambiguities and gauge symmetries.  
In fact, these theories have been widely studied by means of the 
Lagrangian and Hamiltonian formalism. However, such analyses can be 
supplemented and, in many cases improved, by another scheme defined 
for singular systems as it is the case of the Hamilton-Jacobi formalism.

In this paper we construct in a consistent manner a 
Hamilton-Jacobi (HJ) framework for linearly 
acceleration-dependent Lagrangians. In providing this 
scheme we also furnish with a great alternative to 
analyze the phase space constraint structure of this 
type of theories. The HJ framework we are interested in, 
introduced by Carath\'eodory for regular systems~\cite{caratheodory1967} and further strengthened 
from the physical point of view for singular systems by G\"uler and 
many authors~\cite{guler1992a,guler1992b,pimentel1996,pimentel1998,hasan2004,
pimentel2005,pimentel2008}, is esentially based on the variational principle 
named \textit{equivalent Lagrangians method} and the first-order 
partial differential equations theory. This scheme is obtained directly 
from the Lagrangian formalism without going through the usual approximation 
of the canonical transformations of the Hamiltonian formalism, even in 
the case of singular theories, so it becomes an adequate shorcut for 
arriving at the HJ framework. For regular systems we can arrive 
straightforwardly at the commonly known HJ partial differential equation. 
On the contrary, for singular systems constraints arise as necessary 
conditions to ensure the existence of extremes for a given action but, 
without ceasing to be partial differential equations from which the 
characteristic system of equations turn out to depend on several independent 
variables, named parameters, which may be related with the gauge 
information of the systems~\cite{pimentel2014}. The complete set of HJ 
partial differential equations must obey geometric conditions in order 
to guarantee their integrability. These conditions are equivalent to the 
consistency conditions developed in the Dirac-Bergmann approach for 
constrained systems~\cite{Dirac1964,henneaux1992,rothe2010}. In consequence, 
the HJ integrability analysis separates the constraints in involutive 
and non-involutive under an extended Poisson bracket~\cite{pimentel2014,
pimentel2008b}. The existence of non-involutive constraints signals a 
dependence between the parameters of the theory which leads to a 
redefinition of the symplectic structure of the phase space in order to
have control of the right dynamics of the theory. 

In contrast to the Dirac-Bergmann approach, this HJ framework possesses 
robust geometric foundations and does not need support from the split 
of first- and second-class constraints to build the right evolution of 
physical constrained systems as well as to obtain the gauge symmetry 
information. For theories being linear on the accelerations non-involutive 
constraints are always present. To unravel this situation within this 
geometric framework it is mandatory to introduce a generalized bracket 
(GB) structure that leaves the complete set of constraints in involution 
since the so-called Frobenius integrability conditions are 
fulfilled~\cite{pimentel2014}, thus solving the problem of integrability. 
This framework also links the complete set of involutive HJ equations 
with the canonical and gauge symmetries. Regarding this, we find that 
theories that only have non-involutive constraints are characterized by 
having matrices $N_{\mu\nu}$ and $\mathcal{M}_{AB}$ being non-singular,~(\ref{Nij}) 
and~(\ref{MMAB}), respectively, so that the evolution in phase space 
is only generated by the canonical Hamiltonian $\H_0$ under a properly 
defined GB. On the other hand, when matrix $\mathcal{M}_{AB}$ turns out to be 
singular, the evolution in phase space of any observable involves several 
parameters, treated on an equal footing as the time parameter where 
some of them are related to the gauge invariance properties of the Lagrangian 
function behind that theory. At this particular point we highlight the 
uselfulness in taking advantage of the use of the zero-modes 
of the matrix $\mathcal{M}_{AB}$ in order to obtain the Hamiltonians 
that help to identify the reduced physical phase space. Our findings shed 
light on previously overlooked geometrical aspects of HJ approaches for 
affine in accelerations or velocities theories and can serve as the basis 
for quantization schemes by using a WKB approximation. We want to
emphasize here that, despite of being warned that every higher-order 
theory can be written down as a first-order theory by appropriately 
enlarging the configuration space by introducing auxiliary
fields,~\cite{Carinena2003,pimentel2005,pimentel2008}, the point of 
view we wish to implement here is strictly that emerging from the 
higher-order theory. Indeed, in many cases it is more appropriate to 
analyze the original information of the system while maintaining all the 
original symmetries thus avoiding the extension of both the number of 
variables and constraints in the new configuration which, due to the
cumbersome notation involved may hide the geometrical structure of a model.

The paper is organized as follows. In Sect.~\ref{sec1} we 
glance an overview of the kinematical description of the 
affine in acceleration theories and introduce some important
geometrical structures. After established the notation and conventions, 
in Sect.~\ref{sec2} we construct the HJ framework for the affine in 
acceleration theories that yield third-order equations of motion. The HJ 
scheme when second-order equations are present is developed in
Sec.~\ref{sec3} where a couple of possible scenarios is discussed. 
In Sec.~\ref{sec4} our development is illustrated with some 
examples: a Chern-Simons-like theory in $(2+1)$-dimensions,
an harmonic oscillator in 2$D$ and, the RT cosmology in 
extra dimensions. We conclude in Sec.~\ref{sec5} with some remarks 
on our development. Further, here we outline the obtaining of the 
Hamilton principal function $S$ and comment on possible 
extensions of the work. In an Appendix we show the geometric 
role played by the matrix $M_{\mu\nu}$,~(\ref{Mmunu}), in our 
development.

\section{Hamilton-Jacobi formalism for affine in 
acceleration theories}
\label{sec1}

We are interested in physical systems with a finite number 
of degrees of freedom described by Lagrangian functions, 
including only terms linear in accelerations, and described
by the action
\be 
\mathcal{S}[q^\mu] = \int dt\,L(q^\mu,\q^\mu, \qq^\mu, t)
\quad \mu=1,2,\ldots , N;
\label{action1}
\ee
where
\be 
L(q^\mu,\q^\mu, \qq^\mu, t) = \K_\mu(q^\nu,\q^\nu,t)\,\qq^\mu 
- \mathcal{V}(q^\nu,\q^\nu,t).
\label{L0}
\ee
Summation over repeated indices is henceforth understood. 
According to the higher-order derivative theories viewpoint 
the configuration space $\mathcal{C}_{2N}$ is spanned by 
the $N$ coordinates $q^\mu$ and their $N$ velocities $\q^\mu$. 
An overdot stands for the derivative with respect to the 
time parameter $t$ so that $\q^\mu = dq^\mu/dt$ and so on. 
Here, $\K_\mu$ and $\mathcal{V}$ are assumed to be smooth 
functions defined on $\mathcal{C}_{2N}$.

Guided by the Hamilton's principle adapted to actions of the 
form~(\ref{action1}), the optimal trajectory $q^\mu = q^\mu (t)$ 
parametrized by $t$, is obtained by solving the Euler-Lagrange 
(EL) equations of motion (eom) 
\be 
\frac{\partial L}{\partial q^\mu}
- \frac{d}{dt} \left[ \frac{\partial L}{\partial \q^\mu} - 
\frac{d}{dt} \left( \frac{\partial L}{\partial \qq^\mu}
\right) \right] = 0.
\nonumber
\ee
Explicitly, these are given by
\be 
N_{\mu\nu}\dddot{q}^\nu - M_{\mu\nu} \qq^\nu + F_\mu = 0,
\label{eom1}
\ee
where
\beq 
N_{\mu\nu} &:=& \frac{\partial \K_\nu}{\partial \q^\mu} - 
\frac{\partial \K_\mu}{\partial \q^\nu} = - N_{\nu\mu},
\label{Nij}
\\
M_{\mu\nu} &:=&\frac{\partial \K_\nu}{\partial q^\mu} - 
\frac{\partial}{\partial \q^\nu} \left( - \frac{\partial 
\V}{\partial \q^\mu} - \frac{\partial \K_\mu}{\partial q^\rho} 
\q^\rho + N_{\mu \rho} \qq^\rho \right),
\label{Mij}
\\
F_\mu &:=& \frac{\partial}{\partial q^\nu} \left( - \frac{\partial 
\V}{\partial \q^\mu} - \frac{\partial \K_\mu}{\partial q^\rho} 
\q^\rho + N_{\mu\rho} \qq^\rho \right)\q^\nu + \frac{\partial 
\V}{\partial q^\mu}.
\label{Fi}
\eeq
It is worth noting that $N_{\mu\nu} = N_{\mu\nu} (q^\rho,\q^\rho,t)$, 
$M_{\mu\nu} = M_{\mu\nu} (q^\rho,\q^\rho, \qq^\rho,t)$ and $F_\mu = 
F_\mu (q^\rho,\q^\rho, \qq^\rho,t)$. The form~(\ref{eom1}) for 
the eom proves to be fairly useful and allows the theory to 
be better understood in the HJ scheme to be developed.

Following the original \textit{Carath\'eodory's equivalent 
Lagrangians approach}~\cite{caratheodory1967,guler1992a,
guler1992b}, later extended to second-order in derivatives 
theories~\cite{pimentel1996,pimentel1998}, in order to have 
an extreme configuration of the action~(\ref{action1}) 
the necessary conditions are associated to the existence of a 
family of surfaces defined by a \textit{generating function}, 
$S(q^\mu, \q^\mu,t)$, such that it satisfies
\beq 
&&\frac{\partial S}{\partial \q^\mu} = \frac{\partial L}{\partial 
\qq^\mu} =: P_\mu,
\label{eq2a}
\\
& &\frac{\partial S}{\partial q^\mu} = \frac{\partial L}{\partial 
\q^\mu} - \frac{d}{dt} \left( \frac{\partial L}{\partial \qq^\mu}
\right) =: p_\mu,
\label{eq2b}
\\
& & \frac{\partial S}{\partial t} + \frac{\partial S}{\partial 
q^\mu} \q^\mu + \frac{\partial S}{\partial \q^\mu} \qq^\mu - L 
= 0,
\label{eq2c}
\eeq
where, on physical grounds, $P_\mu$ denotes the conjugate momenta 
to the velocities $\q^\mu$ while $p_\mu$ are the conjugate momenta 
to the coordinates $q^\mu$. 

The HJ framework in which we are interested in, emerges 
from~(\ref{eq2a}),~(\ref{eq2b}) and~(\ref{eq2c}) considered as 
partial differential equations (PDE) for $S$. Indeed, for 
non-singular systems it is straightforward to convert~(\ref{eq2c}) 
into a PDE for $S$ by solving for $\qq^\mu$ in terms of $q^\mu$, 
$\q^\mu$ and partial derivatives of $S$ what is obtained by 
appropriately inverting~(\ref{eq2a}). However, for singular physical 
systems this is not possible as the Hessian matrix with elements
$H_{\mu\nu}$ associated to~(\ref{L0}) vanishes identically,
\be 
H_{\mu\nu} := \frac{\partial^2 L}{\partial \qq^\mu \partial 
\qq^\nu} = 0.
\label{eq3}
\ee
This fact defines what is known as an \textit{affine in acceleration 
theory}~\cite{Carinena2003,Udriste2011,Rojas2016}. The rank of the 
Hessian matrix is zero which causes that the manifold $\mathcal{C}_{2N}$ 
is fully spanned by  $R = N - 0 = N$ variables, $\q^\mu$, all of them 
related to the kernel of $H_{\mu\nu}$. In this sense, within the HJ 
scheme we are going to discuss below we have to treat all the generalised 
velocities as free parameters~\cite{guler1992a,guler1992b,pimentel1996}.
Clearly, we can not invert any of the accelerations $\qq^\mu$ in 
favour of the coordinates, the velocities $\q^\mu$ and the momenta 
$P_\mu$ through the partial derivative of the function $S$, fact that 
entails the presence of $N$ constraints given by the definition of 
$P_\mu$ itself. Indeed, as one can infer from (\ref{eq2a}-\ref{eq2c}), 
the constraints form a set of PDE of first-order for $S$ which, in order 
to be integrable, must obey the so-called~\textit{Frobenius 
integrability conditions},~\cite{caratheodory1967,pimentel2014}, 
(see~(\ref{integrabilityG}) below). On the other hand, relying on the 
structure of the momenta $p_\mu$, (see (\ref{pi}) below) we could or not solve for 
the accelerations $\qq^\mu$ in favour of the remaining variables of 
the phase space which could or not give rise to more contraints. Indeed, 
such a dependence rests heavily on the nature of the quantities 
$\K_\mu$ which in turn promote two possible scenarios characterized by
the order of the eom. As a matter of fact, the momenta $P_\mu$ and 
$p_\mu$ written out in full are 
\beq 
P_\mu &=& \K_\mu (t,q^\nu,\q^\nu),
\label{Pi}
\\
p_\mu &=& - \frac{\partial \V}{\partial \q^\mu} - 
\frac{\partial \K_\mu}{\partial q^\nu} \q^\nu + 
N_{\mu\nu}\qq^\nu.
\label{pi}
\eeq
Note the linear dependence on the accelerations in~(\ref{pi}). 
The eom in this framework are written as total differential equations known as \textit{characteristic equations}, \cite{guler1992a,guler1992b}, whose solutions are trajectories 
in dependence of independent variables, $t^\alpha$, in a 
reduced phase space, as we will see below. For the further analysis, due to the existence of several parameters in 
the development, in due course it will necessary to relabel 
the indexes associated with the main geometric quantities. 
For that reason, it is mandatory to discuss sistematically 
the two feasible scenarios.

\section{Theories with third-order equations of motion}
\label{sec2}

According to the Jacobi's theorem in matrix algebra, the determinant of any antisymmetric matrix of odd order has determinant equal to zero. Certainly, by observing 
(\ref{eom1}), not necessarily all coordinates $q^\mu$ will 
obey third-order eom in some sector of the configuration 
space. In order to ensure the existence of independent 
third-order eom we confine ourselves to consider an even 
number of variables $q^\mu$. Hence, we set $N=2n$ implying 
thus that $N_{\mu\nu}$ is non-vanishing and that 
$\det(N_{\mu\nu}) \neq 0$ in this scenario. Under these 
conditions, to determine the time evolution it is needed to consider $6n$ initial conditions for the quantities $q^\mu, 
\q^\mu, \ddot{q}^\mu$ with $\mu=1,2,\ldots, 2n$ at initial 
time $t = t_0$. Additionally, from~(\ref{pi}) it should be 
noted that accelerations can be solved in favour of the 
rest of the phase space variables.

From the result (\ref{eq3}), all the generalized velocities have 
the status of parameters \cite{pimentel2005} so, it is reasonable to 
introduce the notation: $t^0 := t$ 
\be
t^{\mu} := \q^\mu 
\quad  \mbox{and} \qquad
H_{\mu}^P :=  - \frac{\partial L}{\partial \qq^\mu} 
= - \K_\mu (t,q^\nu, t^{\nu}),
\label{def1}
\ee
as well as the set of coordinates $t^I$ in the
following order
\be 
t^I := q^I := (t^0, t^{\mu}),
\qquad \qquad
I 
= 0,1, 2, \ldots, 2n.
\label{def2}
\ee
As a result, relationship~(\ref{eq2a}) reads
\be 
\frac{\partial S}{\partial t^{\mu}} + H_{\mu}^P \left( t, 
t^{\nu}, q^\nu, \frac{\partial S}{\partial t^{\nu}}\right) 
= 0.
\label{eq5}
\ee
In the same spirit, by introducing the~\textit{Hamilton 
function}
\be 
\label{H0}
H_0 := \frac{\partial S}{\partial q^\mu} t^\mu 
+ \frac{\partial S}{\partial t^\mu} \dot{t}^\mu 
- L(t^0,q^\mu, t^\mu,\dot{t}^\mu),
\ee
which does not depend explicitly on $\dot{t}^\mu$, as it may 
be checked straightforwardly, one finds that the 
expression~(\ref{eq2c}) becomes
\be 
\label{eq7}
\frac{\partial S}{\partial t^0} + H_0 \left( t^0,t^{\mu},q^\mu, 
\frac{\partial S}{\partial t^{\mu}}, \frac{\partial 
S}{\partial q^\mu}\right) = 0,
\ee
that is the common expression of the Hamilton-Jacobi equation.
In terms of the original notation, the Hamilton function reads
\be 
H_0 = p_\mu \q^\mu + \V (q^\nu,\q^\nu, t).
\label{H00}
\ee
We can express~(\ref{eq5}) and~(\ref{eq7}) as a unified set of PDE 
for the generating function $S$. To do this, it is useful to assume
that $P_0 := \frac{\partial S}{\partial t^0}$ is conjugate canonical 
momentum to $t^0$. We then find
\be 
\label{eq8}
\frac{\partial S}{\partial t^I} + H_I \left( t^J,q^\mu, 
\frac{\partial S}{\partial q^\mu},\frac{\partial S}{\partial 
t^J}\right) = 0,  \quad
\substack{
I, J = 0,1,2,\ldots, 2n.,
}
\ee
where $H_I := (H_0,H_{\mu}^P)$. In the following, the $2n +1$ relations 
(\ref{eq8}) will be referred to as the \textit{Hamilton-Jacobi partial 
differential equations} (HJPDE). Bearing in mind~(\ref{eq2a}), we can 
also write~(\ref{eq5}) and~(\ref{eq7}) in the form
\beq 
\H_0 &:=& P_0 + H_0 (t^0,t^{\mu}, q^\mu, P_{\mu}, p_\mu)=0,
\label{eq8a}
\\
\H_{\mu}^P &:=& P_{\mu} + H_{\mu}^P (t^0,t^{\nu}, q^\nu)=0,
\label{eq8b}
\eeq
which acquire the compact constrained Hamiltonian fashion
\be 
\H_I (t^J, q^\nu, P_J, p_\nu ) := P_I + H_I (t^J, q^\nu,P_J, 
p_\nu) = 0,
\label{canonC}
\ee
where $\H_I := (\H_0, \H_\mu^P)$ and $P_I := (P_0,P_\mu)$.
These expressions have thus acquired the well-known form of 
canonical Dirac constraints. 
The constraints written in the form~(\ref{canonC}) are also 
referred to as \textit{Hamiltonians} in this HJ 
scheme. 
In a sense, this HJ approach replaces the analysis of the $2n$ 
canonical constraints, $\H_{\mu}^P  = 0$, with 
the analysis of the $(2n+1)$ HJPDE given by relations~(\ref{eq8}). 
Some remarks are in order. 
We do not have a further HJPDE relating the momenta $p_\mu$ 
because from~(\ref{pi}) we observe the linear dependence on 
$\qq^\mu$ which allows us to write the accelerations
in favour of the momenta that can be inserted in the Legendre
transformation defining the canonical Hamiltonian. We shall prove
this in short; in fact, within the Dirac-Bergmann approach for 
constrained systems, this feature signals the presence of 
second-class constraints. On the other hand, the Hamiltonian 
$\H_0$, (\ref{eq8a}), is said to be associated with the 
time parameter $t^0$ while the Hamiltonians $\H_\mu^P$, 
(\ref{eq8b}), are associated with the remaining parameters 
$t^\mu$ related to the velocities of the system.

The equations of motion, known as \textit{characteristic 
equations} (CE), associated to the Hamiltonian set~(\ref{canonC}), are 
given as total differential equations~ \cite{guler1992a,guler1992b}. 
At this initial stage these are given by 
\beq 
dq^\mu &=& \frac{\partial \H_I}{\partial p_\mu} dt^I \qquad
\qquad dq^I = \frac{\partial \H_J}{\partial P_I} dt^J, 
\label{dqs}
\\
dp_\mu &=& -\frac{\partial \H_I}{\partial q^\mu} dt^I \qquad
\quad dP_I = - \frac{\partial \H_J}{\partial q^I} dt^J.
\label{dps}
\eeq
We would like to emphasize that $t^{\mu} = \q^\mu$ have a status of 
independent evolution parameters, on an equal footing to $t$. 
To prove this it is enough to evaluate~(\ref{dqs}) for $\q^\mu$;
indeed, $d\q^\mu = (\partial \H_0/\partial P_\mu)dt^0
+ (\partial \H_\nu^P/\partial P_\mu)dt^\nu = dt^\mu$.
On mathematical grounds, within this HJ formalism it is said that 
$t^I$ are the \textit{independent variables} or \textit{parameters} 
of the theory. In fact, the number of parameters is determined not 
only by the rank of the Hessian matrix but also by the integrability 
conditions. On physical grounds, the parameters encode the local 
symmetries and gauge transformations (see below for details). The 
solution of the first equations in~(\ref{dqs}) and~(\ref{dps}) leads
to a congruence of parametrized curves in the 
configuration space $\mathcal{C}_{2N+1}$, given by $q^\mu = 
q^\mu (t^I)$. In a like manner, the generating function $S(t,q^\mu,
\q^\mu)$ is satisfying
\be 
dS = \frac{\partial S}{\partial t^I} dt^I + \frac{\partial 
S}{\partial q^\mu}dq^\mu  = - H_I \,dt^I + p_\mu dq^\mu,
\label{dS0}
\ee
where~(\ref{eq2b}) and~(\ref{canonC}) have been considered.

For two arbitrary functions $F, G \in \Gamma_{2N+1} := 
T^* \mathcal{C}_{2N+1}$, that is, functions in the extended 
phase space spanned by the variables $ (t^I, q^\mu)$ and 
their conjugate momenta $(P_I, p_\mu)$, we introduce the 
\textit{extended Poisson bracket} (PB)
\be 
\{ F, G \} = \frac{\partial F}{\partial t^I}\frac{\partial 
G}{\partial P_I}  + \frac{\partial F}{\partial q^\mu}\frac{\partial 
G}{\partial p_\mu} -  \frac{\partial F}{\partial P_I}\frac{\partial 
G}{\partial t^I} - \frac{\partial F}{\partial p_\mu}\frac{\partial 
G}{\partial q^\mu}. 
\label{EPB}
\ee
We may therefore express evolution in $\Gamma_{2N+1}$ as follows
\be
dF = \{ F, \H_I \}\,dt^I,
\label{dF}
\ee
where the $t^I$ play the role as parameters of the Hamiltonian flows generated by the Hamiltonians $\H_I$. In passing, the CE~(\ref{dqs}) and~(\ref{dps}) may be obtained from~(\ref{dF}) by evaluating $F$ for any of the phase space variables. In this HJ framework, the dynamical evolution is provided by~(\ref{dF}) which is referred to as the \textit{fundamental differential}.

\subsection{Integrability conditions}
\label{subsubsec:2a}

With the intention of integrating the HJPDE~(\ref{eq8}), it 
is convenient to rely in the method of 
characteristics~\cite{caratheodory1967}. On physical grounds, 
it is completely unclear whether or not  all coordinates 
are relevant parameters of the theory, so it is crucial to 
find a subspace among the parameters $t^I$ where the system 
becomes integrable. Regarding this, the matrix occurring in~(\ref{eom1}) 
\be 
N_{\mu\nu} = \frac{\partial 
\K_\nu}{\partial \q^\mu} - \frac{\partial \K_\mu}{\partial 
\q^\nu} = \{ \H_{\mu}^P, \H_{\nu}^P\},
\label{NAB1}
\ee
plays an important role to unravel under what conditions 
the eom associated with the action~(\ref{action1}) will be 
integrable. 

The complete solution of~(\ref{eq8}) (or~(\ref{canonC})) is 
given by a family of surfaces orthogonal to the characteristic 
curves. In this sense, the fulfillment of the Frobenius 
integrability conditions~\cite{caratheodory1967,pimentel2014}
\be 
\{ \H_I, \H_J \} = C^K _{IJ}\,\H_K,
\label{integrabilityG}
\ee
ensures the existence of such a family where $C^K_{IJ}$ are the 
structure coefficients of the theory. This means that the 
Hamiltonians must close as an algebra. Accordingly, it must be 
imposed that $d\H_0$ and $d\H_{\mu}^P$ are vanishing identically
\be 
d \H_I = 0.
\label{integrability}
\ee
Guided by the aforementioned Jacobi theorem, we have non-involutive 
constraints since we have 
$N_{\mu\nu} \neq 0$ and $\det (N_{\mu\nu}) \neq 0$, that is, a 
regular case. Thence, no new Hamiltonians arise from the realization 
of $d\H_I = 0$ but a relation of dependence between the parameters 
of the theory. Indeed, we have that $dt^{\mu} = - (N^{-1})^{\mu\nu} 
\{ \H_{\nu}^P, \H_0 \}dt^0$ where $(N^{-1})^{\mu\nu}$ denotes the 
inverse matrix of $N_{\mu\nu}$ such that $N_{\nu \rho}(N^{-1})^{\rho 
\mu} = \delta^\mu _\nu$ or $(N^{-1})^{\mu \rho} N_{\rho \nu} = 
\delta^\mu _\nu$. In such a case it is often enough to consider 
that $t^0$ is the independent parameter of the theory. From~(\ref{dF})
we infer now that the evolution of $F \in \Gamma_{2N+1}$ is 
provided by
\be 
\label{dF1}
dF = \{ F, \H_0 \}^*\,dt^0,
\ee
where
\be 
\{ F, G \}^* := \{ F, G\} - \{ F, \H_{\mu}^P \} (N^{-1})^{\mu\nu}
\{ \H_{\nu}^P , G\}.
\label{DB1}
\ee
In this HJ spirit, the remaining variables, $q^\mu$, are referred 
to as \textit{dependent variables}. Note that the $N$ independent
Hamiltonians (\ref{eq8a}) and (\ref{eq8b}) will fix the dynamics
on the phase space in a unique way.

The bracket structure introduced in (\ref{DB1}) is 
referred to as the \textit{generalized bracket} (GB) which 
has all the known properties of the standard Poisson bracket. 
In the present case, this redefines the dynamics by eliminating 
the parameters $t^\mu$ with exception of $t^0$. Accordingly, 
the non-involutive Hamiltonians have been absorbed in the GB. 
As a matter of fact, the GB is closely related to the Dirac bracket 
arising in the Dirac-Bergmann Hamiltonian approach for constrained 
systems~\cite{Dirac1964,henneaux1992,rothe2010}. 
Therefore, under the fundamental differential~(\ref{dF1}), the 
Hamiltonians preserve the condition~(\ref{integrability}) in the 
reduced phase space defined by $\H_{\mu}^P = 0$, and we have 
as a result an integrable set of Hamiltonians.

\subsection{Characteristic equations}
\label{subsubsec:2b}

The characteristic equations may now be computed from~(\ref{dF}). 
First,
\be 
dq^\mu = \{ q^\mu, \H_0 \}^* dt^0 = \q^\mu dt^0,
\label{eom1a}
\ee
which is a trivial identity. Second,
\beq 
dt^{\mu} = d\q^\mu &=& \{ \q^\mu , \H_0 \}^* dt^0,
\nonumber
\\
&=& (N^{-1})^{\mu\nu} \left( p_\nu + \frac{\partial 
\V}{\partial \q^\nu} + \frac{\partial \K_\nu}{\partial 
q^\rho} \q^\rho \right) dt^0,
\label{ch2}
\eeq
provides the accelerations of the mechanical system
as we observe from~(\ref{pi}). Third,
\beq 
dp_\mu &=& \{ p_\mu, \H_0 \}^* dt^0,
\nonumber
\\
&=& -\frac{\partial \V}{\partial q^\mu }dt^0 + 
\frac{\partial \K_{\nu}}{\partial q^\mu} dt^{\nu},
\label{eom1b}
\eeq
represents the equations of motion 
provided by~(\ref{eom1}) by direct substitution of the previous 
characteristic equation and~(\ref{pi}). Finally,
\beq 
dP_\mu &=& \{ P_\mu , \H_0 \}^* dt^0,
\nonumber
\\
&=& \left( - p_\mu - \frac{\partial \V}{\partial \q^\mu} 
\right)dt^0 + \frac{\partial \K_{\nu}}{\partial \q^\mu} d t^{\nu}.
\label{eom1c}
\eeq
This is nothing but the definition of the momenta $p_\mu$, namely
$p_\mu = \partial L/\partial \q^\mu - dP_\mu/dt$, once we 
insert~(\ref{ch2}), in agreement with~(\ref{eq2b}).

Regarding the Hamilton principal function we get
\beq 
dS &=& \{ S, \H_0 \}^* dt^0,
\nonumber
\\
&=& P_0 dt^0 + p_\mu dq^\mu + P_{\mu} dt^{\mu}.
\nonumber
\eeq
In other words,
\be
dS = p_\mu dq^\mu - H_I dt^I,
\label{dS2}
\ee
where the Hamiltonians~(\ref{canonC}) have been introduced. This
expression is in agreement with~(\ref{dS0}). On the other hand,
bearing in mind the nature of the matrix~(\ref{Nij}), we must
recall that the only independent parameter is $t^0$ so that
\be 
dS = p_\mu dq^\mu + P_\mu d\q^\mu - H_0 dt^0.
\label{dS3}
\ee
Some comments are in order. First, it has been inferred that 
the solutions to the characteristic equations, in the complete 
phase space are given by expressions of the form $q^\mu = q^\mu 
(t^I)$, which represent congruences of curves in the $(N+1)$ parametric space, where the $t^I$ play the role of coordinates. From a physical point of view, the actual dynamics of the system is achieved in a reduced phase space and dictated by the fundamental 
differential~(\ref{dF1}). Hence, the solutions to the characteristic 
equations in the physical sector of phase space are given by 
$q^\mu = q^\mu (t^0)$, representing congruences of one-parameter
curves. In comparison with the Ostrogradski-Hamilton analysis for this type 
of theories~\cite{tapia1985,motohashi2015}, we note that we have a 
second-class system as observed from~(\ref{NAB1}) since $N_{\mu\nu} \neq 0$.

\section{Theories with second-order equations of motion}
\label{sec3}

It is readily inferred from~(\ref{eom1}) that to obtain 
second-order equations of motion the matrix $N_{\mu\nu}$ must 
vanish identically. Unlike the previous case, the matrix $M_{\mu\nu}$ 
becomes symmetric. This fact can be proved from~(\ref{Mij}) with 
support with the property $\partial \K_\mu /\partial \q^\nu = 
\partial \K_\nu /\partial \q^\mu$. From the facts that 
now $F_\mu = F_\mu (q^\nu, \q^\nu,t)$ and $M_{\mu\nu} = M_{\mu\nu} (
q^\rho,\q^\rho,t)$, the eom~(\ref{eom1}) specialize to 
Newton-like equations of the form 
\be 
M_{\mu\nu} \qq^\nu = F_\mu \qquad\qquad \mu,\nu = 1,2,\ldots, N;
\ee 
where $M_{\mu\nu}$ can be interpreted as the components of 
a mass-like matrix while the term $F_\mu$ may be intepreted
as a force vector. In fact, the matrix $M_{\mu\nu}$ corresponds
to the Hessian matrix of a first-order equivalent Lagrangian 
$L_d$ (see Appendix~\ref{appen} for details). For this 
particular case, the Hamiltonian constraints~(\ref{eq8a}) and~(\ref{eq8b}) hold. Further, at the initial stage, the 
evolution in phase space is dictated by the fundamental differential~(\ref{dF}). The next step, under the new setup, 
is to test the integrability conditions for the Hamiltonians~(\ref{eq8a}) and~(\ref{eq8b})as we will discuss 
in short. 

\subsection{Integrability conditions}

This particular setting determines a fully constrained system. 
Indeed, to prove this statement we must first mention that 
the expression for the PB (\ref{EPB}) holds and then we must 
to proceed to test the integrability condition for $\H_{\mu}^P$ 
by using~(\ref{dF}) 
\beq 
d \H_{\mu} ^P &=& \{ \H_{\mu} ^P , \H_0 \}\,dt^0
+ \{ \H_{\mu} ^P, \H_{\nu} ^P \} \,dt^{\nu},
\N
\\
&=& - \left( p_\mu + \frac{\partial \V}{\partial t^\mu} 
+ \frac{\partial \K_\mu}{\partial q^\nu}t^\nu \right)dt^0,
\N
\eeq
where we have considered $N_{\mu\nu} = 0$ as we observe 
from~(\ref{NAB1}). Then, we identify new Hamiltonian 
constraints given by
\be
\H_{\mu} ^p := 
p_\mu + \frac{\partial \V}{\partial t^\mu} + \frac{\partial 
\K_\mu}{\partial q^\nu} t^\nu = 0.
\label{eq12}
\ee
The integrability conditions have to be tested with these 
Hamiltonians as well. As before, when separating $t^0$ 
from the remaining parameters $t^{\mu}$ we find
\beq 
d \H_{\mu} ^p &=& \{ \H_{\mu} ^p, \H_0 \} dt^0 + 
\{ \H_{\mu} ^p, \H_{\nu} ^P \} dt^{\nu},
\N
\\
&=& - F_\mu dt^0 + C_{\mu \nu} dt^{\nu} = 0,
\label{eq12a}
\eeq
where we have introduced the antisymmetric matrix
\be 
C_{\mu \nu} := \{ \H_{\mu} ^p, \H_{\nu} ^P \} 
= - C_{\nu \mu},
\ee
and recognize that $\{ \H_0, \H_\mu ^p \} = F_\mu$ as 
defined in~(\ref{Fi}). Clearly, the variations~(\ref{eq12a}) 
do not vanish identically, and in consequence, the idea 
to promote them as new constraints of the theory is 
deceptive. These are mere dependence relationships between 
the parameters of the theory. We then have a complete set 
of Hamiltonians
\beq 
\H_0 &=& P_0 + H_0 ( t^0, t^{\mu}, q^\mu, p_{\mu}) = P_0 
+ p_\mu t^\mu + \V ,
\label{eq10}
\\
\H_{\mu} ^P &=& P_{\mu} + H_{\mu} ^P (t^0, t^{\nu}, q^{\nu}) 
= P_{\mu} - \K_\mu ,
\label{eq11}
\\
\H_{\mu} ^p &=& p_{\mu} + H_{\mu} ^p (t^0, 
t^{\nu}, q^{\nu}, p_\nu)
= p_{\mu} + \frac{\partial \V}{\partial t^{\mu}} + 
\frac{\partial \K_{\mu}}{\partial q^\nu}
t^\nu 
\label{eq12b}
\eeq
Certainly, the Hamiltonians~(\ref{eq10}), (\ref{eq11}) 
and~(\ref{eq12b}) are non-involutive constraints. To satisfy 
the integrability condition it is required to remove the 
non-involutive constraints by redefining the fundamental 
differential trough a generalised bracket structure. To 
perform this, the constraints~(\ref{eq12}) must enter the 
game accompanied of new parameters. In order to further 
analyse the integrability conditions it is necessary to 
introduce a convenient notation and a relabeling of the 
indices. In agreement with the scheme previously outlined in~(\ref{def1}), introducing the quantities
\be
\t^{\bar{\mu}} := q^{\bar{\mu}}
\qquad \text{and} \qquad
H_{\bar{\mu}}^p := - \frac{\partial L}{\partial \q^\mu} + 
\frac{d}{dt}\left( \frac{\partial L}{\partial \ddot{q}^\mu}
\right),
\label{newts}
\ee
with the understanding that $\bar{\mu} = N+1,N+2, \ldots, 
2N$. The $\t^{\bar{\mu}}$ are expected to be in relation 
to the generalized coordinates $q^\mu$. Additionally, we 
introduce the complete set of parameters $t^I$ in the order
\be 
t^I:=(t^0,t^{\mu},\t^{\bar{\mu}})
\qquad \quad
I = 0,1\,\ldots,N,N+1,\ldots, 2N.,
\ee
as well as the notation $H_I := (H_0, H_{\mu}^P, H_{\bar{\mu}}^p)$,
$\H_I = (\H_0, \H_{\mu}^P,\H_{\bar{\mu}}^p)$ and $\mathcal{P}_I 
= (P_0,P_{\mu},p_{\mu})$, respecting that order, with  $I,J = 0,1,2,\ldots, N, N+1, \ldots, 2N$. Surely, we 
can also express the Hamiltonians in a constrained Hamiltonian 
fashion according to~(\ref{canonC})
\be 
\H_I (t^I, \mathcal{P}_I ) = \mathcal{P}_I +
H_I (t^I, \mathcal{P}_I ) = 0,
\label{canonC2}.
\ee
As before, the $2N + 1$ relations~(\ref{canonC2}) represent 
HJPDE. Similarly, as in previous section, it will be said that 
the Hamiltonians $\H_{\bar{\mu}}^p$, (\ref{eq12}), are 
associated with the parameters $\t^{\bar{\mu}}$ that are in 
relation with the coordinates of the physical system.

Now, the evolution of the theory is derived from the 
fundamental differential
\be 
dF = \{ F, \H_0 \} dt^0 + \{ F, \H_\mu ^P \} dt^\mu 
+ \{ F, \H_{\bar{\mu}} ^p \} d\t^{\bar{\mu}},
\label{dF4}
\ee
where the space of parameters has been expanded. At this point
it is worthwhile to remark that the integrability of the system
now should be tested by using~(\ref{dF4}). At this stage, we
should be able to elucidate if the system is integrable either 
in a complete or in a partial manner. By an appropriate relabeling 
of the indices, when using the fundamental differential~(\ref{dF4}) 
as well as separating the time parameter from the remaining ones, 
the condition $d\H_I = \{ \H_I, \H_J \} dt^J = 0$ on the 
Hamiltonians~(\ref{canonC2}) is written as
\be 
d\H_I = \{ \H_I, \H_0 \} dt^0 + \{ \H_I, \H_A \}dt^A.
\label{eq13a}
\ee
Here, $I,J = 0, A$ with $A= \mu,\bar{\mu} = 1,2,\ldots, N, N+ 1, 
\ldots, 2N$ and $\H_A = (\H_\mu ^P, \H_{\bar{\mu}} ^p)$ being the
Hamiltonians organized in that suitable order. Explicitly
\beq 
d\H_0 &=& \{ \H_0, \H_A \} dt^A = 0,
\label{eq14}
\\
d\H_A &=& \{ \H_A, \H_0 \} dt^0 + \mathcal{M}_{AB} dt^B = 0, 
\label{eq15}
\eeq 
where we have introduced the antisymmetric matrix components
$\mathcal{M}_{AB} : = \{ \H_A, \H_B \}$. This is a $2N \times 2N$ 
partitioned matrix 
\be 
\mathcal{M}
=
\begin{pmatrix}
0 & -M_{\mu \bar{\nu}}
\\
M_{\bar{\mu}\nu} & Q_{\bar{\mu}\bar{\nu}}
\end{pmatrix}
\label{MMAB}
\ee
decomposed in terms of the symmetric matrix $M_{\bar{\mu}\nu}$ and 
the antisymmetric one $Q_{\bar{\mu}\bar{\nu}}$ defined as 
\beq 
\{ \H_\mu ^P, \H_{\bar{\nu}}^p \} &=:& - M_{\mu\bar{\nu}}
= - M_{\bar{\nu} \mu},
\label{Mmunu}
\\
\{ \H_{\bar{\mu}} ^p, \H_{\bar{\nu}}^p \} &=:&  Q_{\bar{\mu}\bar{\nu}}
= - Q_{\bar{\nu} \bar{\mu}}.
\label{Qmunu}
\eeq
Notice that~(\ref{Mmunu}) agrees with the expression defining the 
matrix~(\ref{Mij}). If all the parameters are independent, then 
$\{ \H_A, \H_0 \} = 0$ and $\mathcal{M}_{AB} = 0$ are required. 
These conditions lead to the only possible solution of the 
equations of motion and the system becomes integrable. On the 
contrary, it could happen that the Hamiltonians do not obey~(\ref{eq14}) 
and~(\ref{eq15}) so that the fulfillment of the integrability 
conditions leads to assume a linear dependence on the parameters 
$t^I$, leading us to define generalized brackets. For that reason,
the discussion must be addressed on the case where $\mathcal{M}_{AB}$ 
is different from zero. Certainly, 
from~(\ref{eq15}) we get
\be 
\mathcal{M}_{AB} dt^B = - \{ \H_A, \H_0 \} dt^0.
\label{eq16}
\ee
Therefore, at this new stage, the integrability analysis bifurcates.

\begin{itemize}
\item
If $\mathcal{M}_{AB}$ is non-singular then $\det(\mathcal{M}_{AB}) 
\neq 0$. In such a case the inverse matrix $(\mathcal{M}^{-1})^{AB}$ 
exists so that $(\mathcal{M}^{-1})^{AC} \mathcal{M}_{CB} = 
\delta^A{}_B$ or $\mathcal{M}_{AC} (\mathcal{M}^{-1})^{CB} = 
\delta_A{}^B$. Indeed, we will have that
\be 
\mathcal{M}^{-1}
= 
\begin{pmatrix}
(M^{-1}QM^{-1})^{\mu\nu} & (M^{-1})^{\mu\bar{\nu}}
\\
- (M^{-1})^{\bar{\mu} \nu} & 0
\end{pmatrix}.
\ee
The realization of $d\H_I = 0$ leads to consider that $dt^0$  
and $dt^A$ are dependent. We infer from~(\ref{eq16}) that
$t^0 = t$ is the independent parameter of the theory
\be 
dt^A = - \left( \mathcal{M}^{-1} \right)^{AB} \{ \H_B, \H_0 \} dt^0.
\label{eq17}
\ee
In this manner, when substituting~(\ref{eq17}) into~(\ref{eq13a}), 
we observe that the evolution of $F \in \Gamma_{2N+1}$ is
provided by
\be 
dF = \{ F, \H_0 \}^* dt^0,
\label{dF11}
\ee
where
\be 
\{ F, G \}^* := \{ F, G\} - \{ F, \H_A \} (\mathcal{M}^{-1})^{AB}
\{ \H_B, G\}.
\label{DB11}
\ee
This structure helps to redefine the dynamics by eliminating 
the parameters $t^A$ with exception of $t^0$. In passing, we would 
like to mention that the complete set of Hamiltonians $\H_I$ are in 
involution with the bracket structure~(\ref{DB11}). In a sense, 
this set up is similar to one outlined in the previous section.

\item
If $\mathcal{M}_{AB}$ is singular then $\det(\mathcal{M}_{AB}) 
= 0$. This property is related to the existence of a gauge
symmetry in the Lagrangian (\ref{L0}).
The rank of $\mathcal{M}_{AB}$ being, say $R = 2N - r$, 
implies the existence of $r$ left (or right) null eigenvectors 
$\lambda^A_{(\alpha)}$, or \textit{zero-modes}, of $\mathcal{M}_{AB}$ 
such that $\mathcal{M}_{AB} \lambda^B_{(\alpha)} = 0$ where $(\alpha)$
labels the independent zero-modes with $\alpha = 1,2,\ldots, r$. In 
such a case the original configuration manifold $\mathcal{C}_{2N}$, 
not including the time parameter, becomes splitted in two submanifolds: 
$\mathcal{C}_r$ spanned by $r$ coordinates, $t^\alpha$, related to the 
kernel of $\mathcal{M}_{AB}$, and $\mathcal{C}_R$ spanned by $R$ 
coordinates $t^{A'}$, with $A' = r+1,r+2,\ldots,2N$., associated with 
the regular part of $\mathcal{M}_{AB}$. Under these conditions, 
we ensure the existence of a $R\times R$ submatrix, say 
$\mathcal{M}_{A'B'}$, such that $\det(\mathcal{M}_{A'B'}) \neq 0$ 
so that we have an inverse matrix $(\mathcal{M}^{-1})^{A'B'}$ 
satisfying $(\mathcal{M}^{-1})^{A'C'} \mathcal{M}_{C' B'} = 
\delta^{A'} _{B'}$ or $\mathcal{M}_{B'C'} (\mathcal{M}^{-1})^{C'A'} 
= \delta_{B'} ^{A'}$. According to this splitting we set $A = 
\alpha, A' = 1,2,\ldots,r, r+1, \ldots, 2N$.

The expansion of the condition~(\ref{eq16}) reads
\beq 
-\{ \H_A, \H_0 \}dt^0 &=& 
\delta_A ^{A'}\mathcal{M}_{A'B'} dt^{B'} + \delta_A ^\alpha 
\mathcal{M}_{\alpha B'} dt^{B'} 
\N
\\
&+&  \, \delta_A ^{A'} \mathcal{M}_{A'\beta} dt^\beta 
+ \delta_A ^\alpha \mathcal{M}_{\alpha \beta} dt^\beta.
\label{eq18}
\eeq
We must inspect two cases
\begin{itemize}
\item[a)] 
If $A = A'$ it follows that $dt^{A'}$ can be expressed in terms 
of $dt^\alpha$ and $dt^0$. Indeed, from~(\ref{eq18}) we get
\be 
\mathcal{M}_{A'B'} dt^{B'} =: - \{ \H_{A'}, \H_{\alpha'} \} 
dt^{\alpha'},
\N
\ee
where we have introduced the label $\alpha' = 0, 1, 2, \ldots, r$. 
Thence,
\be 
\label{eq19}
dt^{A'} = - (\mathcal{M}^{-1})^{A'B'}\{ \H_{B'}, 
\H_{\alpha'} \} dt^{\alpha'}.
\ee
We have therefore a relation of dependence between the 
parameters $dt^{A'}$ and $dt^{\alpha'}$.

\item[b)]
If $A = \alpha$, from~(\ref{eq18}) again, it is inferred that
\begin{widetext}
\be
- \{ \H_\alpha, \H_0 \} dt^0 - \{ \H_\alpha, \H_\beta \}dt^\beta 
-\{ \H_\alpha,\H_{A'} \} (\mathcal{M}^{-1})^{A'B'} \{ \H_{B'}, 
\H_0 \} dt^0 - \{ \H_\alpha,\H_{A'} \} (\mathcal{M}^{-1})^{A'B'} 
\{ \H_{B'} , \H_\beta \}dt^\beta
\ee
\end{widetext}
where we have inserted~(\ref{eq19}). Based on the linear
independence between $dt^0$ and $dt^\alpha$, it follows
that
\beq 
\{ \H_\alpha , \H_0 \} &=& \{ \H_\alpha,\H_{A'} \} 
(\mathcal{M}^{-1})^{A'B'} \{ \H_{B'} , \H_0 \},
\label{eq20a}
\\
\{ \H_\alpha , \H_\beta \} &=& \{ \H_\alpha,\H_{A'} \} 
(\mathcal{M}^{-1})^{A'B'} \{ \H_{B'} , \H_\beta \},
\label{eq20b}
\eeq
should be considered as conditions that fix the subspace of 
parameters where the system becomes integrable.
\end{itemize}
The aforementioned dependence on the variables,~(\ref{eq19}),
when inserted into~(\ref{eq13a}), allows to determine the 
evolution of $F \in \Gamma_{2N+1}$ as follows
\be 
dF = \{ F, \H_{\alpha'} \}^* dt^{\alpha'},
\label{dF21}
\ee
where
\be 
\{ F, G \}^* := \{ F, G\} - \{ F, \H_{A'} \} (\mathcal{M}^{-1})^{A'B'}
\{ \H_{B'}, G\}.
\label{DB21}
\ee
As already mentioned, the variables $t^{\alpha'}$ are the independent 
parameters of the theory whereas the remaining variables $t^{A'}$ are 
the \textit{dependent variables}. Clearly, under this split 
of the variables $t^A$ into $t^\alpha$ and $t^{A'}$, each of 
them will be in relation to both coordinates and velocities 
of the system.
\end{itemize}
The bracket structure introduced either in~(\ref{DB11}) or 
in~(\ref{DB21}), is also referred to as the \textit{generalized 
bracket}. As a matter of fact, this structure is closely related 
to the Dirac bracket arising in the Dirac-Bergmann approach for 
constrained systems~\cite{Dirac1964,henneaux1992,rothe2010}. 
Therefore, the dynamical evolution of the theory 
depends on $\alpha'$ parameters, $t^{\alpha'}$.

On the other hand, it may happens that the 
constraints $\H_I = 0$ do not satisfy $d\H_I = 0$ identically 
when~(\ref{dF11}) or~(\ref{dF21}) are considered as fundamental 
differentials. In such a case, the integrability condition leads 
us to obtain equations of the form $f(q^\mu,\q^\mu, p_\mu,P_\mu) = 0$ 
which should also be considered as constraints of the system. 
In a like manner, the integrability conditions must be tested 
for $f$, which could also generate more Hamiltonians. Once we 
have found the complete set of involutive Hamiltonians, it is 
mandatory to incorporate them within the HJ framework where 
some of them must be considered as generators of the dynamics.  
This incorporation must be accompanied by the introduction of more 
parameters to the theory, these related to the new constraints that 
generate dynamics, derived from the integrability analysis. 
Thereupon, the space of parameters has been increased where, 
every arbitrary parameter is in relation to the generators of 
the dynamics~\cite{pimentel2008,pimentel2014}. As a result, 
in this new scenario, the final form of the fundamental differential 
reads
\be 
dF = \{ F, \H_{\overline{\alpha}} \}^*\,dt^{\overline{\alpha}},
\label{dF3}
\ee
with the understanding that $t^{\overline{\alpha}}$ denotes the 
complete set of independent parameters where the index 
$\overline{\alpha}$ spans over the entire set of these parameters. 
Thence, the fundamental differential~(\ref{dF3}) must be 
used to obtain the right evolution in the reduced phase space 
through the GB. 

\subsubsection{Integrability analysis based on zero-modes}
\label{subA}

A useful procedure to identify the inversible submatrix
$\mathcal{M}_{A'B'}$ as well as for the search of gauge information
of the physical systems, under the present conditions, is based
on the zero-modes of $\mathcal{M}_{AB}$.  
To develop this, consider a non-singular transformation
$G$ acting on the differentials of the parameters, except $t^0$, 
given by $dt^A = G^A{}_B dt^B$. By expanding the indices 
covering the  ker$(\mathcal{M}_{AB})$ and its complement
subspace, we have $dt^A = G^A{}_\alpha dt^\alpha + G^A{}_{A'} dt^{A'}$.
We can choose now as a suitable basis of the linear transformation 
the zero-modes $\lambda^A_{(\alpha)}$ with $\alpha = 1,2,\ldots, r.,$
and a set of $R$ vectors $\lambda^A_{(A')}$, with $A' = r+1,r+2,\ldots,
2N$, chosen in a way such that they do not depend on the zero-modes 
or on one another and on the condition that $\det (G^A{}_B) \neq 0$. 
Relative to this basis we have that $G^A{}_\alpha = \lambda^A_{(\alpha)}$ and 
$G^A{}_{A'} = \lambda^A_{(A')}$ so that we can build the evolution 
of the system by using the fundamental differential
\be
dF =\{ F, \H_0 \} dt^0 + \{ F, \H_\alpha \} dt^\alpha 
+ \{ F, \H_{A'} \} dt^{A'},
\label{dF5}
\ee
where
\beq 
\H_\alpha &:=&  \H_A \lambda^A_{(\alpha)} = 0,
\label{HHs1}
\\
\H_{A'} &:=& \H_A \lambda^A_{(A')} = 0.
\label{HHs2}
\eeq
From this viewpoint we must now work with the equivalent 
set of $2N+1$ Hamiltonians given by~(\ref{eq8a}),~(\ref{HHs1})
and~(\ref{HHs2}). According to this, the integrability 
of the system should be tested by considering~(\ref{dF5}). 
The integrability condition applied to the projected 
Hamiltonians 
become
\beq 
d\H_0 &=& \{ \H_0,\H_\alpha \} dt^\alpha
+ \{ \H_0, \H_{A'} \} dt^{A'} = 0,
\label{eq20}
\\
d\H_\alpha &=& \{ \H_\alpha, \H_0 \} dt^0 = 0,
\label{eq21}
\\
d\H_{A'} &=& \{ \H_{A'}, \H_0 \} dt^0
+ \mathsf{M}_{A'B'} dt^{B'} = 0,
\label{eq22}
\eeq
where we have defined
\be 
\mathsf{M}_{A'B'} := \mathcal{M}_{AB} \lambda^A_{(A')} 
\lambda^B_{(B')}.
\label{M3}
\ee
This is an antisymmetric non-singular matrix that is everywhere
invertible on the reduced phase space so the existence
of an inverse matrix, say $(\mathsf{M}^{-1})^{A'B'}$,
is ensured. From~(\ref{eq22}) we identify a dependence among 
some of the original parameters. Indeed, we have that
$dt^{A'} = - (\mathsf{M}^{-1})^{A'B'} \{ \H_{B'}, \H_0 \}
dt^0$ so that when inserted into~(\ref{dF5}) it follows 
straightforwardly that the evolution must be given by
\be
dF = \{ F, \H_{\alpha'} \}^* dt^{\alpha'},
\label{dF6}
\ee
where
\be 
\{ F, G\}^* := \{ F, G \} - \{ F, \H_{A'} \} 
(\mathsf{M}^{-1})^{A'B'} \{ \H_{B'}, G \},
\label{GPBf}
\ee
with the understanding that $\H_{\alpha'} := (\H_0, \H_\alpha)$
and $\alpha' = 0,1,2,\ldots,r$. In arriving to the new fundamental
differential~(\ref{dF6}) we have considered that $\{ \H_{A'}, 
\H_\alpha \} = 0$. On the other hand, from~(\ref{eq21}) and
by inserting $dt^{A'}$, deduced from~(\ref{eq22}), 
into~(\ref{eq20}) we have that
\beq 
d\H_0 &=& \{ \H_0, \H_\alpha \} dt^\alpha = 0,
\\
d \H_\alpha &=& \{ \H_\alpha , \H_0 \} dt^0 = 0,
\eeq
are identically vanishing. To prove this, observe that
$\{ \H_0 , \H_\alpha \} dt^0 = \{ \H_0, \H_A \} 
\lambda^A_{(\alpha)} dt^0$; now, from~(\ref{eq16})
we find that $\{ \H_0, \H_\alpha \} dt^0 = - \mathcal{M}_{BA}
\lambda^A_{(\alpha)} dt^0 = 0$ so that it is inferred that
$\{ \H_0, \H_\alpha \} = 0$. Therefore, the fundamental 
differential~(\ref{dF6}) provides the $(r+1)$-parameter 
evolution in the phase space in which the Hamiltonians 
$\H_{\alpha'}$ are the generators. It should be noted that 
Hamiltonians~(\ref{eq8a}),~(\ref{HHs1}) and~(\ref{HHs2}) 
become involutive under the GB given
by~(\ref{GPBf}) so that the theory becomes integrable.
We would like to mention that this alternative approach 
is suitable for systems exhibiting a neatly geometric 
structure as it is the case, for instance, for extended 
objects evolving in an ambient spacetime \cite{defo1995}.

\subsection{Characteristic equations}
\label{subsec:3}

The characteristic equations that arise in the regular case 
governed by the fundamental differential~(\ref{dF11}), are 
similar to those obtained in the previous section. On the 
other hand, for the singular case, having at our disposal 
the GB~(\ref{GPBf}), the dynamics of the system is provided 
by the fundamental differential~(\ref{dF6}) by evaluating 
$F$ for the phase space variables. In this spirit, it will be convenient to write the 
characteristic equations in terms of the components of the 
zero-modes $\lambda^A_{(\alpha)}$. First, for $F=q^{\bar{\mu}}$
we have
\beq 
dq^{\bar{\mu}} &=& \{ q^{\bar{\mu}}, \H_0 \}^* dt^0 
+ \{ q^{\bar{\mu}}, \H_\alpha \}^* dt^\alpha,
\N
\\
&=& 
\left[
\q^{{\mu}} - \lambda^{\bar{\mu}}_{(A')}
(\mathsf{M}^{-1})^{A'B'} \{ \H_{B'}, \H_0 \} \right] dt^0,
\N
\\
&+& \lambda^{\bar{\mu}}_{(\alpha)} dt^\alpha,
\label{char1}
\eeq
where we have considered that $\{ \H_{B'}, \H_\alpha \} = 0$. 
Second, for $F = \q^\mu$ we obtain
\beq 
d \dot{q}^{\mu} &=& - (\mathsf{M}^{-1})^{A'B'}\frac{\partial 
\H_{A'}}{\partial P_\mu} \{ \H_{B'}, \H_0 \} dt^0 + \lambda^\mu_{(\alpha)} 
dt^\alpha.
\label{char2}
\eeq
For $F= P_\mu$, we get 
\beq 
dP_\mu &=& \left[ - p_{\bar{\mu}} - \frac{\partial \V}{\partial \q^\mu} +
(\mathsf{M}^{-1})^{A'B'}\frac{\partial \H_{A'}}{\partial \q^\mu}
\{ \H_{B'}, \H_0 \} \right]dt^0
\N
\\
&-& 
\frac{\partial \H_\alpha}{\partial \q^\mu} dt^\alpha.
\label{char3}
\eeq
Finally, for $F=p_{\bar{\mu}}$
\beq 
dp_{\bar{\mu}} 
&=& \left[ - \frac{\partial \V}{\partial q^{\bar{\mu}}} +  
(\mathsf{M}^{-1})^{A'B'}\frac{\partial \H_{A'}}{\partial q^{\bar{\mu}}}
\{ \H_{B'}, \H_0 \} \right] dt^0 
\N
\\
&-& \frac{\partial \H_\alpha}{\partial q^{\bar{\mu}}} dt^\alpha.
\label{char4}
\eeq
From this viewpoint, the characteristic equations provide on the 
one hand the time evolution whereas, on the other hand, provide 
the canonical and gauge transformations by analysing the 
evolution of the system at fixed time $\delta t^0$ along the 
remaining parameters. To correctly reproduce the equations 
of motion, it will be necessary to choose appropriate parameters 
$t^\alpha$. This is so since when calculating the characteristic 
equations from the original form~(\ref{dF4}), we can observe 
that the differentials $dt^\mu$ are arbitrary so, under the 
conditions in this scenario, the right dynamics in the physical 
phase space is fixed by choosing some values for the 
parameters. 
 
Regarding the Hamilton principal function, $S= S(q^{\bar{\mu}},
\dot{q}^\mu,t^0)$, we have that
\beq 
dS &=& \{ S, \H_0 \}^* dt^0 + \{ S, \H_\alpha \}^* dt^\alpha,
\N
\\
&=& \frac{\partial S}{\partial q^{\bar{\mu}}} \dot{q}^{\bar{\mu}}
dt^0 + \left( \frac{\partial S}{\partial q^{\bar{\mu}}} \lambda^{\bar{\mu}}_{(A')} 
+  \frac{\partial S}{\partial \dot{q}^{\mu}} \lambda^{\mu}_{(A')}  \right) dt^{A'}
\N
\\
&+& \left( \frac{\partial S}{\partial q^{\bar{\mu}}} \lambda^{\bar{\mu}}_{(\alpha)} 
+  \frac{\partial S}{\partial \dot{q}^{\mu}} \lambda^{\mu}_{(\alpha)}  \right) dt^{\alpha}.
\N
\eeq
Taking into account~(\ref{eq2a}),~(\ref{eq2b}) and~(\ref{eq2c})
projected along the $\lambda^A_{(\alpha,A')}$ vectors, and collected
into $\mathcal{P}_I = (P_0, p_\alpha, p_{A'}, P_\alpha, P_{A'})$
we have
\be 
dS = - H_{\alpha'} dt^{\alpha'} + p_{A'} dt^{A'} + P_{A'} dt^{A'},
\label{dSp}
\ee
where we have considered~(\ref{canonC2}). 
In summary,~(\ref{char1}-\ref{dSp}) define a reduced phase space
with coordinates $(q^{A'},\q^{A'},p_{A'},P_{A'})$.
As we already mentioned, given some initial conditions, the 
solution to the characteristic equations will be dynamical 
trajectories restricted to the variables $q^{A'}$ whose 
parametric equations will be of the form $q^{A'} = q^{A'}(t,t^\alpha)$.

\subsection{Generator of gauge symmetries}
\label{subsec:2b}

Once the complete set of involutive Hamiltonians, 
$\H_{\overline{\alpha}}$, satisfying  $\{ \H_{\overline{\alpha}}, 
\H_{\overline{\beta}} \}^* = C^{\overline{\gamma}}_{\overline{\alpha} 
\overline{\beta}}\, \H_{\overline{\gamma}}$, has been obtained, 
these must be considered as generators of infinitesimal canonical 
transformations of the form, 
\cite{pimentel2014} 
\be 
\delta z^{\bar{A}} = \{ z^{\bar{A}}, \H_{\overline{\alpha}} \}^*\,\delta 
t^{\overline{\alpha}},
\label{flows1}
\ee
where $z^{\bar{A}} = (q^{A'},\mathcal{P}_{A'})$. These are referred 
to as the \textit{characteristic flows} of the system. Here, $\delta 
t^{\overline{\alpha}} := \bar{t}^{\,\,\overline{\alpha}} - 
t^{\overline{\alpha}} = \delta t^{\overline{\alpha}} 
(z^{\bar{A}})$. In particular, when these transformations are taken 
at constant time, $\delta t^0 = 0$, the expression~(\ref{flows1}) 
defines a special class of transformations
\be 
\delta z^{\bar{A}} = \{ z^{\bar{A}}, \H_{\dot{\alpha}} \}^*\,\delta t^{\dot{\alpha}},
\label{flows2}
\ee
which, by observing that they remain in the reduced phase space, 
$T^* \mathcal{C}_P$, form the so-called infinitesimal contact 
transformations in the spirit of the constrained Hamiltonian 
framework by Dirac~\cite{Dirac1964}. In this sense, the
transformations (\ref{flows2}) do not alter the physical states 
of the system. Thus, $t^{\dot{\alpha}}$ denotes the set of all independent 
parameters where $t^0$ is excluded. Clearly, transformations~(\ref{flows2}) 
are generated by 
\be 
G := \H_{\dot{\alpha}} \,\delta t^{\dot{\alpha}},
\label{G}
\ee
so that~(\ref{flows2}) is equivalent to
\be 
\delta_G z^{\bar{A}} = \{ z^{\bar{A}}, G \}^*.
\label{flows3}
\ee
Thus, $\delta_G z^{\bar{A}}$ is the specialization of~(\ref{flows1}) to  
$T^* \mathcal{C}_r$ 
where $G$ is the generating function of the infinitesimal 
canonical transformation. In the spirit of the theory of gauge 
fields, transformations~(\ref{flows3}) stand for the gauge transformations of the theory, \cite{pimentel2014}.

\section{Applications}
\label{sec4}

Having disposal of the general aspects of our development
we can now turn to consider some examples that
illustrate how the above schemes work.

\subsection{Galilean invariant $(2+1)$-dim model with a 
Chern-Simons-like term}

Consider the effective Lagrangian of a non-relativistic 
system~\cite{Lukierski1997}
\be 
L (q^\mu, \q^\mu, \qq^\mu, t) = -k \epsilon_{\mu\nu} \q^\mu \qq^\nu
+ \frac{m}{2} \q^2 \quad \quad \mu,\nu = 1,2.,
\ee
where $k$ is a parameter of the theory, $m$ is a constant
and $\epsilon_{\mu\nu}$ is the Levi-Civita antisymmetric metric
with $\epsilon_{12} = 1$.
This one-particle model with second-order derivatives, describes
a free motion in the $D=2$ space with non-commutating coordinates 
and internal structure described by oscillator modes with negative
energies.
Once we identify the basic structures $\mathcal{K}_\mu = k 
\epsilon_{\mu\nu}\q^\nu$ and $\mathcal{V} = -(m/2)\q^2$, from (\ref{Nij}), (\ref{Mij}) and (\ref{Fi}), it is straightforward to compute
\be 
N_{\mu\nu} = -2k \epsilon_{\mu\nu},
\qquad M_{\mu\nu} = - m \delta_{\mu\nu} \qquad \text{and} \qquad
F_\mu = 0,
\ee
respectively. Hence, from (\ref{eom1}) the eom are given by
\be 
\label{eom3}
2k \epsilon_{\mu\nu} \dddot{q}^\nu - m \delta_{\mu\nu} \qq^\nu = 0,
\ee
which are of third-order in the derivatives. On the other hand,
from (\ref{Pi}) and (\ref{pi}), the momenta for this theory 
reads
\beq 
P_\mu &=& k\epsilon_{\mu\nu}\q^\nu, 
\label{Pmu}
\\
p_\mu &=& m \q_\mu - 2k \epsilon_{\mu\nu} \qq^\nu.
\label{pmu}
\eeq
The canonical Hamiltonian $H_0 = p_\mu \q^\nu + P_\mu \qq^\mu - L$
can be readily obtained
\be 
H_0 = p_\mu \q^\mu - \frac{m}{2} \q^2.
\ee
Now, from (\ref{eq8}) (or (\ref{canonC})), we have the HJPDE
\beq 
\mathcal{H}_0 &=& \frac{\partial S}{\partial t}
+ \frac{\partial S}{\partial q^\mu} t^\mu 
- \frac{m}{2}\delta_{\mu\nu}t^\mu t^\nu,
\\
\mathcal{H}_{\mu}^P &=& \frac{\partial S}{\partial t^{\mu}}
- k \epsilon_{\mu\nu} t^{\nu} = 0,
\eeq
or,
\beq
\mathcal{H}_0 &=& P_0 + p_\mu \q^\mu - \frac{m}{2}
\delta_{\mu\nu}t^\mu t^\nu,
\\
\mathcal{H}_{\mu}^P &=& P_{\mu} - k\epsilon_{\mu\nu}t^{\nu} = 0,
\eeq
in a constrained Hamiltonian fashion.
By noticing that the inverse matrix of (\ref{Nij}), for 
the present case, is given by $(N^{-1})^{\mu\nu} = (1/2k)
\epsilon^{\mu\nu}$, the GB (\ref{DB1}) reads
\be 
\{ F, G \}^* = \{ F, G \} - \frac{1}{2k} \{ F, \mathcal{H}_{\mu}^P
\} \epsilon^{\mu \nu}  \{ \mathcal{H}_{\nu}^P , G \},
\label{GBk}
\ee
where $F$ and $G$ are phase space functions.
Defining $Q^\mu := 2\q^\mu$, $Q:=2/k$ and $K := k/2$, we are 
able to find the non-vanishing fundamental generalized brackets 
of the theory 
\be
\begin{array}{ll}
\,\,\,\{ q^\mu,p_\nu \}^{*} = \delta^\mu{}_\nu & \,\,\quad
\,\,\{ Q^\mu , Q^\nu \}^{*} = Q \,\epsilon^{\mu\nu},
\\
\{ Q^\mu , P_\nu \}^{*} = \delta^\mu{}_\nu & \qquad
\{ P_\mu , P_\nu \}^{*} = K \epsilon_{\mu\nu}.
\end{array}
\ee
We thus find that $(q^\mu,p_\nu)$ and $(Q^\mu, P_\nu)$ are the canonical 
pairs of the theory under the GB structure. In passing, we observe 
that under the GB (\ref{GBk}) the coordinates $Q^\mu$ are 
non-commutative. Regarding the characteristic equations, by considering 
the results of~\ref{sec3} we have that~(\ref{eom1a}) is a mere identity. In the 
same spirit,~(\ref{ch2}) leads to $dt^\mu = d \q^\mu = (1/2k)\epsilon^{\mu\nu}(p_\nu 
- m \q_\nu) dt^0$ which is in agreement with~(\ref{pmu}). On the 
other hand~(\ref{eom1b}) yields to $dP_\mu = -(p_\mu -m \q_\mu)dt^0
- k \epsilon_{\mu\nu}dt^\nu$ which is also in agreement with~(\ref{Pmu}).
Finally,~(\ref{eom1c}) leads to $dp_\mu = 0$. This last fact plays a double
duty. On one hand, this leads to the equation of motion~(\ref{eom3})
once we insert the expression~(\ref{pmu}). On the other hand, this yields
the fact that the momenta $p_\mu$ are constant.

It is worthwhile to mention that in \cite{Lukierski1997}
this system has been analysed using the equivalent canonical
Hamiltonian given by
\be 
H_0 = - \frac{m}{2k^2}P^2 + \frac{1}{k} \epsilon^{\mu\nu}P_\mu p_\nu,
\ee
where the authors focused on the quantum properties of the 
system.

\subsection{Harmonic oscillator in 2D}

A case where the matrix (\ref{MMAB}) is regular is provided
by the Lagrangian~\cite{Goldstein1980}
\be
L (q^\mu, \q^\mu, \qq^\mu,t) = - \frac{m}{2}q_\mu \qq^\mu
- \frac{k}{2} q^2 \qquad \mu,\nu = 1,2. 
\label{lagHarmonic}
\ee
where $m$ and $k$ are constants. This is a non-standard
way to study an isotropic harmonic oscillator as we will
see shortly. We readily identify that $\K_\mu
= -(m/2)\delta_{\mu\nu} q^\nu$ and $\V = (k/2)\delta_{\mu\nu}
q^\mu q^\nu$. From~(\ref{Nij}),~(\ref{Mij}) and~(\ref{Fi}) we 
find that
\be 
N_{\mu\nu} = 0 \qquad M_{\mu\nu} = - m \,\delta_{\mu\nu}
\quad \text{and} \quad 
F_\mu = k \,q_\mu.
\ee
Consequently, from~(\ref{eom1}) we have the second-order 
equations of motion
\be 
m \, \qq^\mu + k q^\mu = 0,
\ee
which is nothing but the well known Hooke's law. Regarding the 
momenta, from~(\ref{Pi}) and~(\ref{pi}), we get 
\be
P_\mu = - \frac{m}{2} q_\mu 
\qquad \text{and} \qquad
p_\mu = \frac{m}{2} \q_\mu.
\label{momentaHar}
\ee
For this pedagogical case, the corresponding Legendre 
transformation yields
\be 
H_0 = p_\mu \q^\mu + \frac{k}{2} q^2.
\ee

On account of the expressions~(\ref{eq10}),~(\ref{eq11}) 
and~(\ref{eq12b}), the Hamiltonians of the theory are 
\beq
\H_0 &=& P_0 + p_\mu \q^\mu + \frac{k}{2} q^2 = 0,
\\
\H_\mu^P &=& P_\mu + \frac{m}{2} q_\mu = 0,
\\
\H_\mu^p &=& p_\mu - \frac{m}{2} \q_\mu = 0.
\eeq
These expressions, together with (\ref{EPB}), allow us to 
quickly determine that the matrix $Q_{\mu\nu}$, (\ref{Qmunu}), 
vanishes. Thus, the partitioned matrix (\ref{MMAB}) turns out 
to be non-singular. This, and its inverse, are given by
\be
\mathcal{M} = m
\begin{pmatrix}
0 & 0 & 1 & 0
\\
0 & 0 & 0 & 1
\\
-1 & 0 & 0 & 0
\\
0 & -1 & 0 & 0
\end{pmatrix}
\quad
\mathcal{M}^{-1} =
\frac{1}{m} 
\begin{pmatrix}
0 & 0 & -1 & 0
\\
0 & 0 & 0 & -1
\\
1 & 0 & 0 & 0
\\
0 & 1 & 0 & 0
\end{pmatrix},
\ee
respectively. Under these conditions, $t^0 = \tau$ is the 
only independent parameter of the model and the evolution 
in the phase space is provided by the fundamental differential 
given by~(\ref{dF11}) where, on account of~(\ref{DB11}), the 
GB is given by
\beq 
\{F,G\}^* &=& \{F,G\} + \frac{1}{m} \{ F, \H_\mu^P\}
\delta^{\mu\nu} \{ \H_\nu^p , G\}
\N
\\
&& \qquad \quad \,- \frac{1}{m}\{F,\H_\mu^p\}\delta^{\mu\nu} 
\{ \H_\nu^P,G\}.
\eeq
Having at our disposal this GB and by defining $\pi_\mu :=
2p_\mu$ and $\Pi_\mu := 2P_\mu$ we find that the non-zero
fundamental generalized brackets are 
\be
\begin{array}{ll}
\{ q^\mu, \pi_\nu \}^* = \delta^\mu{}_\nu
& \qquad \{ q^\mu, \q^\nu \}^* = \frac{1}{m}\delta^{\mu\nu}
\\
\{ \q^\mu, \Pi_\nu \}^* = \delta^\mu{}_\nu
& \qquad \{ \pi_\mu, \Pi_\nu \}^* = m \delta_{\mu\nu}.
\end{array}
\label{FGB2}
\ee
Apparently $(q^\mu,\pi_\nu)$ and $(\q^\mu,\Pi_\nu)$ are the 
canonical pairs of the theory but this fact is deceptive. 
This is definitely an example of the  fact that the theory 
described by (\ref{lagHarmonic}) needs a boundary term in 
order to restore the original properties of a harmonic oscillator 
without the need to describe it in terms of second-order terms.
Regarding the characteristic equations, when using (\ref{dF11})
and (\ref{FGB2}), all the dynamical and geometrical information
is correctly reproduced, as stated at the beginning 
of \ref{subsec:3}.

\subsection{Geodetic brane cosmology with a cosmological 
constant}

Consider the cosmological effective Lagrangian for a geodetic
brane-like universe governed by the Regge-Teitelboim (RT)
model~\cite{RT1975,Tapia1989,Maia1989,Pavsic2001,Rojas2009,
Davidson1998,Davidson2003,Paston2010,Paston2012}
\be 
L (a,\dot{t},\dot{a},\ddot{t},\ddot{a},\tau)
= \frac{a^2\dot{t}}{N^3} \left(  \dot{t} \ddot{a}
-  \dot{a}\ddot{t} \right) + \frac{a}{N} \Upsilon,
\label{lagRT1}
\ee
where $\Upsilon := t^2 - N^2 a^2 \bar{\Lambda}^2$.
Here, and in what follows, $\bar{\Lambda}^2 := \Lambda/3\alpha$
where $\Lambda$ and $\alpha$ are constants and $N:= 
\sqrt{\dot{t}^2 - \dot{a}^2}$ represents the \textit{lapse}
function that commonly appears when we perform an ADM 
decomposition of the RT model. Further, a dot stands for 
derivative with respect to the time parameter $\tau$ where $q^\mu 
= (t(\tau),a(\tau))$. This model is invariant under 
reparameterizations of the coordinates (for more details 
see \cite{Rojas2009,Rojas2020}). 
For this theory we readily identify $\mathcal{K}_\mu = a^2 
\dot{t} (- \dot{a}, \dot{t})/N^3$ and $\mathcal{V} = - 
(a/N) \Upsilon$ with $\mu=1,2$ and 
$\bar{\mu} = 3,4$. From these, we compute the basic 
structures~(\ref{Nij}),~(\ref{Mij}) and~(\ref{Fi}): $N_{\mu\nu} 
= 0$,
\be 
(M_{\mu\bar{\nu}}) = \frac{a\Phi}{N^5}
\begin{pmatrix}
\dot{a}^2 & - \dot{t} \dot{a}
\\
-  \dot{t} \dot{a} & \dot{t}^2
\end{pmatrix},
\qquad
(Q_{\bar{\mu}\bar{\nu}}) = \frac{\dot{t}\Theta}{N^3}
\begin{pmatrix}
0 & - 1
\\
1 & 0
\end{pmatrix},
\label{MyQ}
\ee
and
\be 
(F_{\bar{\mu}}) = -\frac{\dot{a}\Theta}{N^3}
\begin{pmatrix}
-\dot{a}
&
 \dot{t}
\end{pmatrix},
\label{FiRT}
\ee
where
\beq 
\Theta &:=& \dot{t}^2 - 3 N^2 a^2  \bar{\Lambda}^2,
\label{Theta}
\\
\Phi &:=& 3 \dot{t}^2 - N^2 a^2  \bar{\Lambda}^2.
\label{Phi}
\eeq
Note that $\det(M_{\mu\bar{\nu}}) = 0$ where its rank is $r=1$. 
Contrary to this fact, we have that $\det(Q_{\bar{\mu}\bar{\nu}}) 
\neq 0$. 
By inserting~(\ref{MyQ}) and~(\ref{FiRT}) into~(\ref{eom1}) 
we find a solely equation of motion
\be 
\frac{d}{d\tau}\left( \frac{\dot{a}}{\dot{t}}\right) =
- \frac{N^2}{a \dot{t}} \frac{\Theta}{\Phi},
\label{eomRT}
\ee
which is of second-order in derivatives in the variables
$t(\tau)$ and $a(\tau)$.

The momenta of the theory, from (\ref{Pi}) and (\ref{pi}),
are
\be
(P_\mu) = \frac{a^2 \dot{t}}{N^3}
\begin{pmatrix}
-\dot{a}
&
 \dot{t}
\end{pmatrix}
\qquad \text{and} \qquad
(p_{\bar{\mu}}) = - \frac{a\Upsilon}{N^3}
\begin{pmatrix}
-\dot{t}
&
 \dot{a}
\end{pmatrix}.
\label{Piypi1}
\ee
In this spirit, the corresponding Legendre 
transformation yields
\be 
H_0 = p_t \dot{t} + p_a \dot{a} - \frac{a}{N} \Upsilon.
\label{H0RT1}
\ee
On physical grounds, the HJ analysis gets more convenient 
when using the projector approach based on zero-modes. The 
partitioned matrix~(\ref{MMAB}) is given by
\be 
\mathcal{M} = \frac{\F}{N^4}
\begin{pmatrix}
0 & 0 & -\dot{a}^2 & \dot{t}\dot{a} 
\\
0 & 0 & \dot{t}\dot{a} & - \dot{t}^2
\\
\dot{a}^2 & - \dot{t}\dot{a} & 0 & - N^2\frac{\T}{\F}
\\
- \dot{t}\dot{a} & \dot{t}^2 & N^2 \frac{\T}{\F} & 0
\end{pmatrix},
\label{MABRT}
\ee
where $\T := (\dot{t}/N)\Theta$ and $\F := (a/N)\Phi$
with $\Theta$ and $\Phi$ defined in (\ref{Theta}) and (\ref{Phi}),
respectively. The rank of this matrix, 
being $R=2$, signals the presence of two zero-modes. Indeed, 
guided by the notation introduced in Sect.~\ref{sec3} and in the 
obtaining of the vectors $\lambda^A_{(\alpha,A')}$ as depicted 
in~\ref{subA} we have that the vectors
\be 
\lambda_{(\alpha)} =
\begin{cases}
\lambda_{(1)} = \frac{1}{2}
\begin{pmatrix}
\dot{t}
\\
\dot{a}
\\
0 
\\
0
\end{pmatrix}
\\
\lambda_{(2)} =
\begin{pmatrix}
\T \dot{a}
\\
\T \dot{t}
\\
- \F\dot{t}
\\
- \F \dot{a}
\end{pmatrix}
\end{cases}
\quad 
\lambda_{(A')} =
\begin{cases}
\lambda_{(3)} =
\begin{pmatrix}
\dot{a}
\\
\dot{t}
\\
0 
\\
0
\end{pmatrix}
\\
\lambda_{(4)} =
\begin{pmatrix}
0
\\
0
\\
\dot{a}
\\
\dot{t}
\end{pmatrix}
\end{cases},
\label{modesRT}
\ee
span the kernel of~(\ref{MABRT}) and its complement subspace, 
respectively, where  $\alpha = 1,2$ and $A'= 3,4$. Bearing 
in mind that $t^0 = \tau$, the original HJPDE for the 
present case are given by
\beq 
\mathcal{H}_0 &=& \frac{\partial S}{\partial \tau}
+ t^{\bar{\mu}} \frac{\partial S}{\partial t^{\bar{\mu}}}
- \frac{a}{N} \Upsilon = 0,
\N
\\
\mathcal{H}_{\mu}^P &=& \frac{\partial S}{\partial t^\mu}
- \frac{a^2 \dot{t}}{N^2} n_\mu = 0,
\N
\\
\mathcal{H}_{\bar{\mu}}^p &=& \frac{\partial S}{\partial 
t^{\bar{\mu}}} + \frac{a \Upsilon}{N^3} \q_{\bar{\mu}} = 0,
\N
\eeq
where we have introduced the vectors
\be 
\q^\mu = 
\begin{pmatrix}
\dot{t}
\\
\dot{a}
\end{pmatrix}
\quad \text{and} \quad
n^\mu = \frac{1}{N}
\begin{pmatrix}
\dot{a}
\\
\dot{t}
\end{pmatrix},
\ee
which are orthogonal in the sense that $\eta_{\mu\nu}\q^\mu 
n^\nu = 0$ where $(\eta_{\mu\nu}) = \text{diag} (-1,1)$. 
In this sense note that $\eta_{\mu\nu}n^\mu n^\nu = 1$.
In fact, these represent both the time vector field and the 
normal vector to the brane-like universe. By projecting the original 
Hamiltonians along~(\ref{modesRT}), as dictated by~(\ref{HHs1}) 
and~(\ref{HHs2}), we get
\be 
\H_\alpha = 
\begin{cases}
\H_1 = \frac{1}{2} C_1,
\\
\H_2 = \T\, C_2 - \F \,C_3,
\end{cases}
\quad
\H_{A'} =
\begin{cases}
\H_3 = C_2,
\\
\H_4 = C_4,
\end{cases}
\label{hamsRT}
\ee
where
\beq 
C_1 &:=& P_t \dot{t} + P_a \dot{a} = 0,
\\
C_2 &:=& P_t \dot{a} + P_a \dot{t} - \frac{a^2 \dot{t}}{N} = 0,
\\
C_3 &:=& p_t \dot{t} + p_a \dot{a} - \frac{a}{N}\Upsilon = 0,
\\
C_4 &:=& p_t \dot{a} + p_a \dot{t} = 0.
\eeq
Note that $C_3 = \H_0$ which identically vanishes as a consequence
of the reparameterization invariance of the model. In the 
representation~(\ref{hamsRT}) the Hamiltonians split into involutive, 
$\H_\alpha$, and non-involutive, $\H_{A'}$, ones. The extended 
Poisson algebra among the $C_i$, with $i=1,2,3,4$ is as follows
\be 
\begin{array}{ll}
\{ C_1, C_2 \} = 0 & \qquad \{ C_2, C_3 \} = - C_4 
\\
\{ C_1, C_3 \} = - C_3 & \qquad \{ C_2, C_4 \} = - C_3 - \F,
\\
\{ C_1, C_4 \} = - C_4 & \qquad \{ C_3, C_4 \} = - \T. 
\end{array}
\label{poissonC}
\ee
Clearly, the $C_i$ determine a non-involutive set of phase
space functions. In this sense, the Hamiltonians (\ref{hamsRT})
obey the extended Poisson algebra
\be
\begin{array}{ll}
 \{ \H_1, \H_2 \} = - \H_2 & \quad
\{ \H_2, \H_3 \} = \frac{6a\dot{t}\dot{a}}{N\F}\,\H_2
- \F\, \H_4
\\
 & \qquad \qquad \qquad + \frac{3\dot{a}}{N\F} (\F \Upsilon 
 - 2a \dot{t}\T)\H_3
 \\
\{ \H_1,\H_3 \} = 0 & \quad \{ \H_2, \H_4 \} = 
- \frac{2\dot{t}^3}{N\F}\,\H_2
\\
 & \qquad \qquad + \frac{2\dot{t}^2}{N\F}(\dot{t}\T 
 - 3N^2 a \bar{\Lambda}^2\F)\H_3
\\
\{ \H_1, \H_4 \} = - \frac{1}{2}\,\H_4 & \quad 
\{ \H_3, \H_4 \} = \frac{1}{\F}\,\H_2 - \frac{\T}{\F}\,\H_3
- \F,
\end{array}
\ee
what verifies that $\H_{A'} = (\H_3,\H_4)$ are non-involutive
Hamiltonians. From~(\ref{M3}) and~(\ref{modesRT}) we can find 
the regular submatrix embedded in the partitioned 
matrix~(\ref{MABRT}) as well as its inverse
\be 
\mathsf{M} = \F 
\begin{pmatrix}
0 & -1
\\
1 & 0
\end{pmatrix}
\qquad \text{and}\qquad
\mathsf{M}^{-1} =
\frac{1}{\F}
\begin{pmatrix}
0 & 1
\\
-1 & 0
\end{pmatrix}
\label{M4}
\ee
respectively. Under these conditions the complete dynamics 
of the theory is dictated by the fundamental differential 
(\ref{dF6}) where the corresponding GB is given by
\beq
\{ F, G \}^* &=& \{ F, G \} - \frac{1}{\F}
 \{ F, \H_3 \} \{ \H_4, G \}
\N
\\ 
&& \qquad \,\quad + \frac{1}{\F} \{ F, \H_4\}\{ \H_3, G \}.
\label{GBrt}
\eeq
Once we have reduced the phase space by constructing the 
appropriate GB, (\ref{GBrt}), we can extract physical information
of the theory. In this spirit, the non-vanishing fundamental
GB between the phase space variables are
\begin{widetext}
\be
\begin{array}{lll}
\{ t, \dot{t} \}^* = - \frac{\dot{a}^2}{\F}
& \quad \{ a, \dot{t} \}^* = - \frac{\dot{t}\dot{a}}{\F}
& \quad \{ \dot{t}, P_a \}^* = - \frac{\dot{a}p_t}{\F}
\\
\{ t, \dot{a} \}^* = - \frac{\dot{t} \dot{a}}{\F}
& \quad \{ a, \dot{a} \}^* = - \frac{\dot{t}^2}{\F}
& \quad \{ \dot{a}, P_t \}^* = - \frac{\dot{t}p_a}{\F}
\\
\{ t, p_t \}^* = 1
& \quad \{ a, p_a \}^* = 1 - \frac{2\dot{t}^2}{\F}
& \quad \{ \dot{a}, P_a \}^* = 1 - \frac{\dot{t}p_t}{\F}
\\
\{ t, p_a \}^* = - \frac{2\dot{t}\dot{a}}{\F}
& \quad \{ a, P_t \}^* =  \frac{\dot{t}P_a - \dot{a}P_t}{\F}
& \quad \{ p_a, P_t \}^* = - \frac{2\dot{t}p_a}{\Phi}
\\
\{ t, P_t \}^* =  \frac{\dot{a}P_a}{\F} + \frac{a^2
\dot{a}^3}{N^3\F}
& \quad \{ a, P_a \}^* =  \frac{2\dot{t}P_t}{\F}
& \quad \{ p_a, P_a \}^* = - \frac{2\dot{t}p_t}{\Phi}
\\
\{ t, P_a \}^* =  \frac{2\dot{a}P_t}{\F} 
& \quad \{ \dot{t}, P_t \}^* =  1 - \frac{\dot{a}p_a}{\F}
& \quad \{ P_t, P_a \}^* =  \frac{a p_t}{\Phi}
\end{array}
\ee
\end{widetext}
Clearly, the pair $(t, p_t)$ is the unique canonical one 
so we have the presence of only a physical degree of 
freedom in this theory. It is worthwhile to mention that
the Hamiltonians $\H_\alpha$ obey  a truncated Virasoro
algebra as discussed in \cite{Biswajit2014,Qmodified2014,
Rojas2020}.

\subsubsection{Characteristic equations}

Guided by (\ref{dF6}), (\ref{GBrt}) and (\ref{char1}-\ref{char4}) 
the evolution along the complete set of parameters is given 
as follows. First,
\be 
\begin{array}{ll}
dt = \dot{t} \,d\tau - \F\,\dot{t}\,dt^2 
& \quad d\dot{t} = - \frac{\T}{\F}\dot{a}\,d\tau 
+ \frac{1}{2}\dot{t}\,dt^1 + \T\,\dot{a}\,dt^2,
\\
da = \dot{a}\,d\tau - \F\,\dot{a}\,dt^2
& \quad d\dot{a} = - \frac{\T}{\F}\dot{t}\,d\tau 
+ \frac{1}{2}\dot{a}\,dt^1 + \T\,\dot{t}\,dt^2,
\end{array}
\label{char5}
\ee
where we have used the values of the zero-modes given by (\ref{modesRT}).
Second,
\be
dp_t = 0 \qquad \qquad dp_a = \left( - \frac{2a\dot{t}}{N} 
\frac{\T}{\F} + \frac{\T}{\dot{t}}\right)d\tau
- \frac{a}{\dot{t}}\T \Upsilon\,dt^2
\label{char6}
\ee
and  
\be
\begin{aligned}
dP_t & = \left[ -p_t - \frac{a\dot{t}}{N^3} \left( 
\dot{t}^2 - N^2 (2 - a^2 \bar{\Lambda}^2)\right)
\right.
\\
& \left. + \frac{\T}{\F}\frac{a^2}{N^3}(\dot{t}^2 + \dot{a}^2)
\right] d\tau
\\
& +  \frac{a^2 \dot{t}\dot{a}}{2N^3} \,dt^1 - \frac{a^2 
\dot{t}}{N^4}\left[ (\dot{t}^2 + \dot{a}^2)
\Theta - 2 \dot{a}^2 \Phi \right] dt^2,
\\
dP_a & = \left[ -p_a + \frac{a\dot{a}}{N^3} \left( 
\dot{t}^2 + N^2 a^2 \bar{\Lambda}^2)\right)
- \frac{\T}{\F}\frac{2a^2\dot{t}\dot{a}}{N^3}
\right] d\tau
\\
& -  \frac{a^2 \dot{t}^2 }{2N^3} \,dt^1 + \frac{2 a^2 \dot{t}^2 
\dot{a}}{N^4} ( \Theta - \Phi ) \,dt^2.
\end{aligned}
\label{char7}
\ee
By choosing $dt^1 = (2/\dot{t}\F) (\F\,d\dot{t} + \T\,da)$ and 
$dt^2 \neq dt^2 (\tau)$ we are able to obtain the eom (\ref{eomRT})
as well as to recover the definition for the momenta $P_\mu$. 
It is worthwhile to mention that the momenta $p_t$ is a constant 
of motion which is a result of the invariance under reparameterizations
of this brane theory. Indeed, this corresponds to the conserved
bulk energy conjugate to the time coordinate $t(\tau)$. 

\subsubsection{Gauge transformations}

As stated above, the Hamiltonians $\H_{\alpha'}$ are 
responsible for generating the complete dynamics along
the directions of the parameters $t^{\alpha'}$. Regarding
the gauge transformations for this theory, the generator 
function $G$ can be constructed from (\ref{G}),
\be 
G = \H_\alpha \delta t^\alpha \qquad\quad
\alpha = 1,2;
\ee
where these transformations are taken at constant time,
$\delta t^0$. It is therefore inferred, from (\ref{flows3}),
that the gauge transformations are given by
\be 
\delta_G z^{\bar{A}} = \{ z^{\bar{A}}, G \}^* = 
\{ z^{\bar{A}}, \H_\alpha \}^* \delta t^\alpha.
\ee
From (\ref{hamsRT}) we have that
\be 
\begin{array}{ll}
\delta_G \,t  = - \F\dot{t}\,\delta t^2
& \qquad \,\,
\delta_G \,\dot{t} = \frac{1}{2}\dot{t}\,\delta t^1
+ \T\,\dot{a}\,\delta t^2,
\\
\delta_G \,a = - \F\dot{a}\,\delta t^2
& \qquad \,\,
\delta_G \,\dot{a} = \frac{1}{2}\dot{a}\,\delta t^1
+ \T\,\dot{t}\,\delta t^2.
\end{array}
\ee
Similarly, regarding the momenta, we obtain
\be 
\delta_G\,p_t  = 0
\qquad \qquad
\delta_G\,p_a = - \frac{a \Upsilon \Theta}{N}\,\delta t^2,
\ee
and
\be 
\begin{aligned}
\delta_G \,P_t &= - \frac{1}{2}P_t \,\delta t^1
- \left[ 2\T\,P_a - 2 \F \,p_t
- \frac{a^2}{N}\T
\right. 
\\
& \left. \qquad \qquad
 + \frac{2a\dot{t}}{N} \F (1 - a^2
\bar{\Lambda}^2) \right]\delta t^2
\\
\delta_G \,P_a &= - \frac{1}{2}P_a \,\delta t^1
- 2\left( \T\,P_t -  \F \,p_a
+ \frac{a\dot{a}}{N}\F\, a^2 \bar{\Lambda}^2 \right)\delta t^2
\end{aligned}
\ee
Therefore, by considering the definitions $\epsilon_2
:= \F\,\delta t^2$ and $2 \epsilon_1:= \delta t^1$, one 
is able to show that $\delta L = 0$ whenever $\epsilon_2 (\tau)$
is vanishing at the extrema located at $\tau = \tau_1$
and $\tau = \tau_2$. This brane model has recently been studied 
also using an HJ approach but, using auxiliary variables 
where the number of Hamiltonians grows enormously.
\cite{Rojas2020}. Regarding the gauge symmetries, it was shown 
in \cite{Rojas2020} that, by considering $\epsilon_1$ and $\epsilon_2$,
$\delta_G t$ and $\delta_G a$ reflect the presence of the 
invariance under reparameterizations of the model (\ref{lagRT1})
while $\delta_G \dot{t}$ and $\delta_G \dot{a}$ reflect the 
presence of an inverse Lorentz-like transformation in the 
velocities.

\section{Concluding remarks}
\label{sec5}

In this paper we have analyzed the integrability of the 
linearly acceleration-dependent theories whithin the
Hamilton-Jacobi framework for second-order singular systems. 
Our construction entails mainly two different scenarios 
according to the nature of the equations of motion. In both 
cases it was shown the presence of a GB structure as a 
consequence that the original theory, from the beginning, 
contains non-involutive constraints. Unlike the Dirac-Bergmann approach for constrained systems, whithin the HJ framework 
it is not mandatory to split the constraints into first- and second-class although they are closely related.

On the other hand, when the partitioned 
matrix~(\ref{MMAB}) turns out to be singular and when the 
involutive constraints can be identified,  these determine 
the so-called characteristic flows~(\ref{flows2}) which are 
in connection with the gauge transformations that leave the 
action~(\ref{action1}) invariant. Along this line, in order 
to illustrate our HJ development some theories were considered
where the obtained results are in agreement with previous 
analyzes. Although is out of the scope of the paper, the 
obtaining of the Hamilton principal function is an element 
that need to be explored in detail. 
Referring to this, despite that the abstract procedure is 
clear  regarding the calculation process, one should always 
be cautious. On the one hand, one may to use the PDE 
techniques in order to solve the HJPDE or, on the other 
hand, when one knows the form of the coordinates and 
momenta as functions of the parameters it is possible to 
perform the integration of the corresponding characteristic equation which in general is rather involved. Attempts in 
this direction have been done in \cite{hasan2014}. 
The analysis here achieved has been carried out, for 
simplicity, for systems with a finite number of degrees of 
freedom. For field theories we believe that it is possible 
to extend our analysis as long as we are careful with the functional analysis. In a sense, our development pave the 
way to be applied to relativistic second-order geometric 
systems characterized by a linear dependence on the 
accelerations as in the case of the so-called Lovelock-like 
brane models that pursue explanation of cosmological acceleration   phenomena~\cite{lovelock1971,Rojas2013,Rojas2016b,Rojas2015} 
where we will explore the WKB approximation for this type of cosmological theories. This subject will be reported elsewhere.

\begin{acknowledgments}
The authors thank  Alberto Molgado for the critical 
reading and enlightening remarks that improved the work and 
also thank Rub\'en Cordero for suggestions and comments 
received. ER thanks the Departamento de F\'\i sica de 
la Escuela Superior de F\'\i sica y Matem\'aticas del I.P.N, M\'exico, where part of this work was developed during a 
sabbatical leave.
ER acknowledges encouragment from ProDeP-M\'exico, 
CA-UV-320: \'Algebra, Geometr\'\i a y Gravitaci\'on. Also, 
ER thanks the partial support from Sistema Nacional de Investigadores, M\'exico. AAS acknowledges support from a 
CONACyT-M\'exico doctoral fellowship.
\end{acknowledgments}

\appendix

\section{On the matrix $M_{\mu\nu}$}
\label{appen}

The matrix $M_{\mu\nu}$ contains important geometric information 
so it is not just a notational resource. This is nothing but the Hessian 
matrix associated to a first-order Lagrangian, $L_d$, where a surface 
term, $L_s$, in the Lagrangian~(\ref{L0}) is identified. To prove this, 
suppose that we are able to write the Lagrangian~(\ref{L0}) in terms 
of a dynamical Lagrangian and a surface Lagrangian as follows
\be
L (q^\mu,\q^\mu,\qq^\mu,t) = L_d (q^\mu,\q^\mu,t) + L_s (q^\mu,\q^\mu, 
\qq^\mu,t),
\label{L1}
\ee
where 
\be
L_s :=  \frac{d h(q^\mu,\q^\mu,t)}{dt},
\label{Ls}
\ee
for a smooth function $h(q^\mu,\q^\mu,t)$. From the Ostrogradski-Hamilton
viewpoint, the momenta are given by
\beq 
P_\mu &=& \frac{\partial L}{\partial \qq^\mu} = 
\frac{\partial L_s}{\partial \qq^\mu},
\N
\\
&=& \mathcal{K}_\mu,
\label{Pi2}
\\
p_\mu &=& \frac{\partial L}{\partial \q^\mu} - \frac{d}{dt} \left( 
\frac{\partial L}{\partial \qq^\mu}\right)
= \frac{\partial L_d}{\partial \q^\mu} + \frac{\partial L_s}{\partial 
\q^\mu} - \frac{d}{dt} \left( \frac{\partial L_s}{\partial \qq^\mu}
\right),
\N
\\
&=& - \frac{\partial \V}{\partial \q^\mu} - \frac{\partial 
\K_\mu}{\partial q^\nu}\q^\nu. 
\label{pi2}
\eeq
Clearly, the momenta conjugate to the coordinates can be written
in terms of the corresponding momenta arising from $L_d$ and $L_s$,
\be 
p_\mu = p_{d\,\mu} + p_{s\,\mu}.
\N
\ee
On the one hand we have that
\be 
\frac{\partial p_\mu}{\partial \q^\nu} = \frac{\partial p_{d\,\mu}}{\partial \q^\nu}
+ \frac{\partial p_{s\,\mu}}{\partial \q^\nu}
=  W_{\mu\nu} + \frac{\partial p_{s\,\mu}}{\partial \q^\nu},
\label{id1}
\ee
where $W_{\mu\nu} := \partial p_{d\,\mu}/\partial \q^\nu$ is 
the Hessian matrix associated with the Lagrangian $L_d$. On the other hand, we have that
\be
\frac{\partial p_\mu}{\partial \q^\nu} = \frac{\partial}{\partial \q^\nu}
\left( - \frac{\partial \V}{\partial \q^\mu} - \frac{\partial 
\K_\mu}{\partial q^\rho}\q^\rho \right) = \frac{\partial \K_\nu}{\partial 
q^\mu} - M_{\mu\nu},
\label{id2}
\ee
where we have substituted (\ref{Mij}) defining the matrix 
$M_{\mu\nu}$. Hence, by considering (\ref{Pi2}), from (\ref{id1}) and
(\ref{id2}) it follows that $M_{\mu\nu} = - W_{\mu\nu}$,
provided that
\be 
\frac{\partial \K_\nu}{\partial q^\mu} = \frac{\partial p_{s\,\mu}}{\partial 
\q^\nu}.
\label{id4}
\ee
We have therefore proved that, when a surface term is existing
in the Lagrangian (\ref{L0}), the matrix $M_{\mu\nu}$ is nothing
but the Hessian matrix associated to the dynamical Lagrangian
$L_d$ whenever the relationship (\ref{id4}) holds.





\end{document}